\documentclass[sigconf]{acmart}
\AtBeginDocument{%
  \providecommand\BibTeX{{%
    \normalfont B\kern-0.5em{\scshape i\kern-0.25em b}\kern-0.8em\TeX}}}

\copyrightyear{2023}
\acmYear{2023}

\usepackage{booktabs}
\usepackage{tcolorbox}
\usepackage{multirow}
\usepackage{bbding}
\usepackage{utfsym}
\usepackage{ulem}
\usepackage{xcolor}
\usepackage{url}
\usepackage{multirow}
\usepackage{tcolorbox}
\usepackage{graphicx}
\usepackage{enumitem}
\usepackage{wasysym}
\usepackage{xspace}
\usepackage{algorithm}
\usepackage{algorithmic}
\usepackage{wasysym}

\begin{document}

\title{\method: Utilizing Existing Code to Enhance Code Generation}

\author{Jia Li \male}
\affiliation{%
  \institution{Peking University}
  \city{Beijing}
  \country{China}
}
\email{lijia@stu.pku.edu.cn}

\author{Yunfei Zhao}
\affiliation{%
  \institution{Peking University}
  \city{Beijing}
  \country{China}}
\email{zhaoyunfei@pku.edu.cn}

\author{Yongmin Li}
\affiliation{%
  \institution{Peking University}
  \city{Beijing}
  \country{China}}
\email{liyongmin@pku.edu.cn}

\author{Ge Li}
\affiliation{%
  \institution{Peking University}
  \city{Beijing}
  \country{China}}
\email{lige@pku.edu.cn}

\author{Zhi Jin}
\affiliation{%
  \institution{Peking University}
  \city{Beijing}
  \country{China}}
\email{zhijin@pku.edu.cn}

\begin{abstract}
Large Language Models (LLMs) have shown great success in code generation. LLMs take as the input a prompt and output the code. A key question is how to make prompts (\ie \textit{Prompting Techniques}).
Existing prompting techniques are designed for natural language generation and have low accuracy in code generation.

In this paper, we propose a new prompting technique named \method.
Our motivation is that code generation meets two unique challenges (\ie requirement understanding and code implementation). \method contains two novel mechanisms (\ie guided code generation and example retrieval) to solve these challenges.
(1) Guided code generation asks LLMs first to analyze requirements and output an intermediate preliminary (\eg test cases). The preliminary is used to clarify requirements and tell LLMs \textit{``what to write''}.
(2) Example retrieval selects similar programs as examples in prompts, which provide lots of relevant content (\eg algorithms, APIs) and teach LLMs \textit{``how to write''}.
We apply \method to three LLMs (\eg Codex) and evaluate it on three public benchmarks using the Pass@$k$. 
Results show that \method can significantly improve the performance of LLMs on code generation.
\textbf{(1) In terms of Pass@1, \method outperforms the state-of-the-art baseline by up to 56.4\% in MBPP, 70.7\% in MBJP, and 88.4\% in MBJSP.}
(2) \method is effective in LLMs with different sizes (\ie 6B to 13B) and different languages (\ie Python, Java, and JavaScript).
(3) Human evaluation shows human developers prefer programs from \method.
\end{abstract}



\def\eg{\textit{e.g.,} }
\def\ie{\textit{i.e.,} }
\def\method{{\sc AceCoder}\xspace}

\renewcommand{\algorithmicrequire}{ \textbf{Inputs:}} 
\renewcommand{\algorithmicensure}{ \textbf{Outputs:}} 


\maketitle

\section{Introduction}
\label{sec:introduction}

Code generation aims to automatically generate the source code based on a natural language requirement \cite{SkCoder,CodeEditor,TiP}. Recently, Large Language Models (LLMs) have achieved state-of-the-art (SOTA) results on code generation \cite{Codex,AlphaCode,CodeGen,InCoder,CodeGeeX}. LLMs do not require fine-tuning and take a prompt as input. A prompt consists of several examples (\eg $<$requirement, code pairs$>$) and a new requirement. LLMs learn code generation from examples and analogously generate code for the new requirement.

The performance of LLMs strongly relies on the prompt surface \cite{PromptStudy}. Thus, how to design prompts (\ie prompting techniques) is a popular topic. Existing prompting techniques (\eg few-shot prompting \cite{GPT-3} and chain-of-thought prompting \cite{Chain-of-thought}) are designed for natural language generation and have low accuracy in code generation.
For example, Codex with few-shot prompting only achieves 37.2\% Pass@1 on a real-world benchmark - HumanEval \cite{Codex}.
Thus, it is necessary to explore more advanced prompting techniques for code generation.

\textbf{In this paper, we propose a novel prompting technique specialized in code generation, named \method.} It significantly improves the performance of LLMs in code generation. Our motivation is that code generation aims to build a mapping from natural language requirements to source code. There are two unique challenges in this mapping, \ie requirement understanding and code implementation.
\method proposes two novel mechanisms to alleviate two challenges. The details of \method are shown as follows.

\textbf{Challenge 1: Requirement Understanding.} 
Understanding requirements is the starting point of code generation. In real-world programming problems, the requirement may be a brief purpose without specific details. For example, a requirement from a real-world benchmark - MBPP \cite{MBPP} is \texttt{write a function to check if the triangle is isosceles or not}. Before writing code, we need to analyze the requirement and determine specific details, \eg input-output formats, and possible exceptions.

\textbf{Novelty 1: Guided Code Generation.}
To alleviate this challenge, we propose \textit{guided code generation}. Our motivation is that human developers often use some software artifacts to assist in analyzing requirements. For example, in test-driven development \cite{Test_Driven_Development}, developers clarify requirements by designing test cases. It forces developers to think about details of requirements, \eg input-output formats and boundary values. These test cases exactly define the requirement and tell developers \textit{what to write}.

To implement the above process, we design a special prompt consisting of triple examples (\ie $<$requirement, preliminary, code$>$). A preliminary is a specific software artifact (\eg test cases, APIs) for clarifying the requirement. Given a new requirement, based on the prompt, LLMs first output a preliminary and then generate code based on the preliminary. We illustrate the guided code generation in Section \ref{sec:motivating_ex} and describe the details in Section \ref{sec:approach:prompt_construction}.

\textbf{Challenge 2: Code Implementation.} 
After understanding the requirement, how to implement the source code using a programming language is challenging. It requires LLMs to master related grammar, algorithms, and libraries. 
Even for human developers, it is difficult to write an exactly correct program from scratch.

\textbf{Novelty 2: Example Retrieval.}
To solve the above challenge, we propose \textbf{example retrieval}. It is inspired by the human developers' code reuse. In real-world scenarios, given a new requirement, developers often search for similar programs. They learn programming skills (\eg APIs) or directly reuse relevant content from similar programs \cite{Study_Code_Reuse}.

Specifically, we use a \textit{retriever} to search for programs with similar requirements (\eg Top-20). Considering the maximum input length of LLMs is limited (\eg 1024 tokens), the number of examples in a prompt is also limited, such as three examples. Thus, we further design a \textit{selector} to select a set of programs from retrieved results as examples. The selector will filter out redundant programs and pick informative examples.
Then, examples are inserted into prompts and teach LLMs how to implement code. 
We illustrate the example retrieval in Section \ref{sec:motivating_ex} and describe the details in Section \ref{sec:approach:example_retrieval}.

In conclusion, given a requirement, \method generates a program in three steps:
\begin{itemize}[leftmargin=*]
    \item \textbf{Example retrieval.} It uses a \textit{retriever} and a \textit{selector} to find similar programs as examples, \ie <requirement, code> pairs.
    \item \textbf{Prompt construction.} It uses an \textit{analyzer} to convert retrieved examples into <requirement, preliminary, code> triples. Then, it concatenates triple examples with the input requirement together to construct a prompt.  
    \item \textbf{Code generation.} It feeds the prompt into LLMs. By learning from examples, LLMs first output an intermediate preliminary and then generate code for the input requirement.
\end{itemize}

We apply \method to three representative LLMs, \ie CodeGeeX \cite{CodeGeeX}, CodeGen \cite{CodeGen}, and InCoder \cite{InCoder}. We conduct extensive experiments on three popular code generation benchmarks, \ie MBPP (Python) \cite{MBPP}, MBJP (Java) \cite{MBXP}, and MBJSP (JavaScript) \cite{MBXP}. We employ Pass@$k$ ($k=1, 3, 5$) to measure the performance of different approaches. 
We obtain some findings from experimental results. 
\textbf{(1) \method significantly outperforms existing prompting techniques.} In terms of Pass@1, \method outperforms the SOTA baseline - few-shot prompting by up to 56.4\% in MBPP, 70.7\% in MBJP, and 88.4\% in MBJSP. The improvements prove the superiority of \method in code generation. 
\textbf{(2) \method substantially outperforms retrieval-based models.} In terms of Pass@1, \method outperforms the SOTA retrieval-based baseline by up to 13.1\% in MBPP, 23.44\% in MBJP, and 15.8\% in MBJSP.
\textbf{(3) \method is effective in LLMs of different sizes.} We apply \method to three LLMs, which scale from 6B to 13B. In terms of Pass@1, \method improves CodeGeeX-13B by up to 88.4\%, CodeGen-6B by up to 65.5\%, and InCoder-6B by up to 57.5\%. 
\textbf{(4) Human evaluation shows that human developers prefer programs generated by \method.} Results show that \method outperforms the SOTA baseline in multiple aspects, including correctness, code smell, and maintainability.
\textbf{(5) We explore the contributions of different modules and discuss different designs for \method.} Results show that three modules are all necessary and our designs for three modules are superior to multiple alternates.

We summarize our contributions in this paper as follows.
\begin{itemize}[leftmargin=*]
    \item We propose a novel prompting technique named \method, for improving the performance of LLMs in code generation.
    \item \method contains two novel techniques (\ie guided code generation and example retrieval) to alleviate two challenges (\ie requirement understanding and code implementation) in code generation, respectively.
    \item We apply \method in three LLMs and conduct extensive experiments on three public benchmarks. Qualitative and quantitative experiments show that \method significantly outperforms the SOTA baselines (\eg chain-of-thought prompting, few-shot prompting).
\end{itemize}

\section{Motivating Examples}
\label{sec:motivating_ex}

\begin{figure}[t]
\centering
\includegraphics[width=0.85\linewidth]{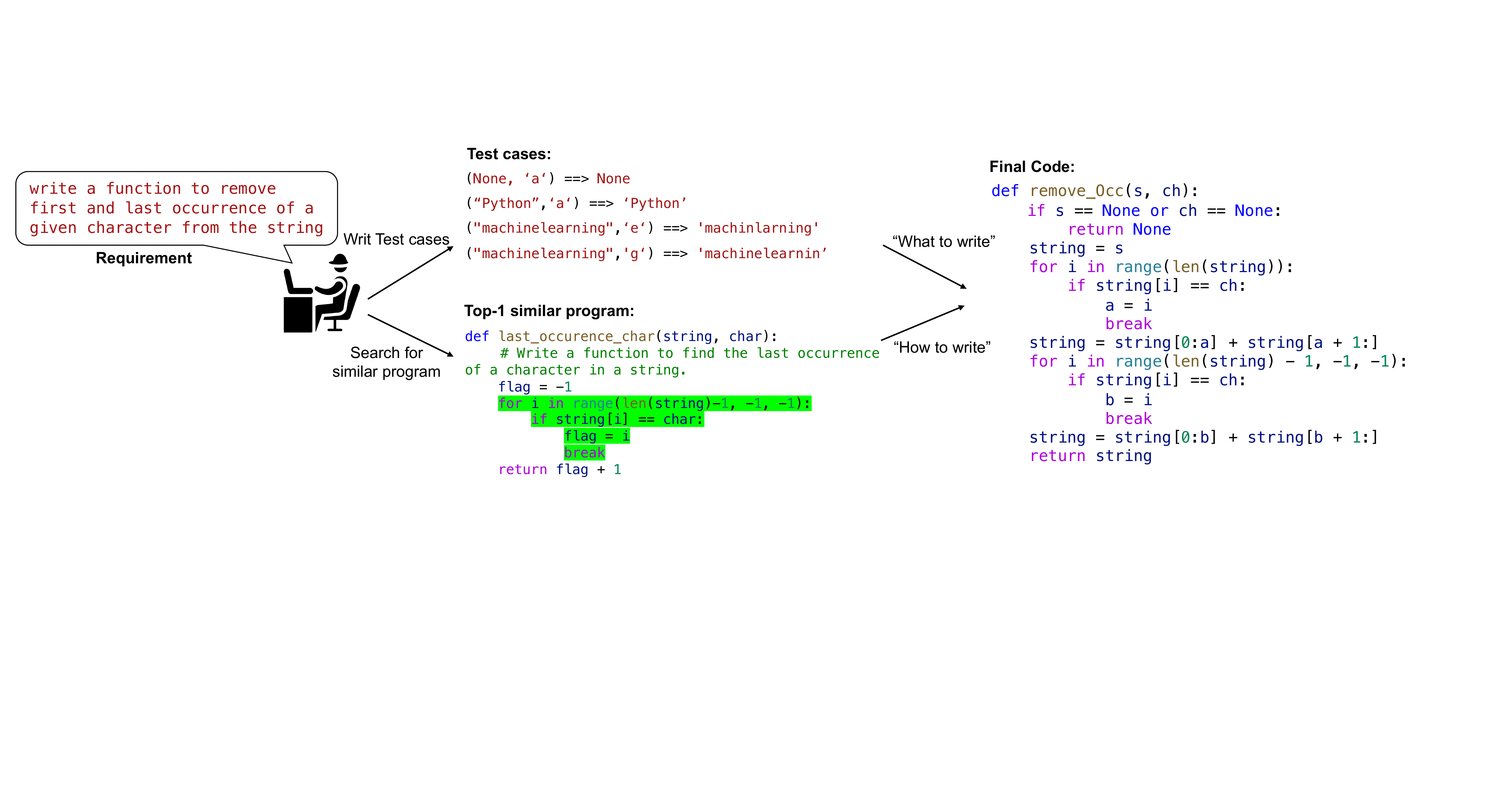}
\vspace{-0.2cm}
\caption{A motivating example of guided code generation.}
\label{fig:motivating_ex1}
\end{figure}

In this section, we explain our motivations by some real cases.

\textbf{Requirement Understanding $\rightarrow$ Guided Code Generation.} Figure \ref{fig:motivating_ex1} (a) and (b) show a requirement from a real-world benchmark \cite{MBPP} and its unit test for evaluation, respectively. 
We select Codex as the base model. Figure \ref{fig:motivating_ex1} (c) shows a program generated by few-shot prompting. The program fails, as it ignores some essential scenarios in the requirement, such as \texttt{ch} appearing multiple times in \texttt{s}. It shows that comprehensively understanding the requirement is crucial to write correct programs. 

Thus, we propose guided code generation, which asks LLMs to first analyze the requirement and then generate code. Figure \ref{fig:motivating_ex1} (d) shows a program generated by \method. We consider test cases as the intermediate preliminary. We can see that the generated test cases cover multiple scenarios, \eg boundary inputs \texttt{("test", "e")}. They further clarify the requirement and benefit the following code implementation. 
Based on the test cases, \method generates a correct program, which considers three scenarios and gives solutions respectively. The example shows that our guided code generation can help LLMs to analyze requirements and improve the logical correctness of code.

\begin{figure}[t]
\centering
\includegraphics[width=0.85\linewidth]{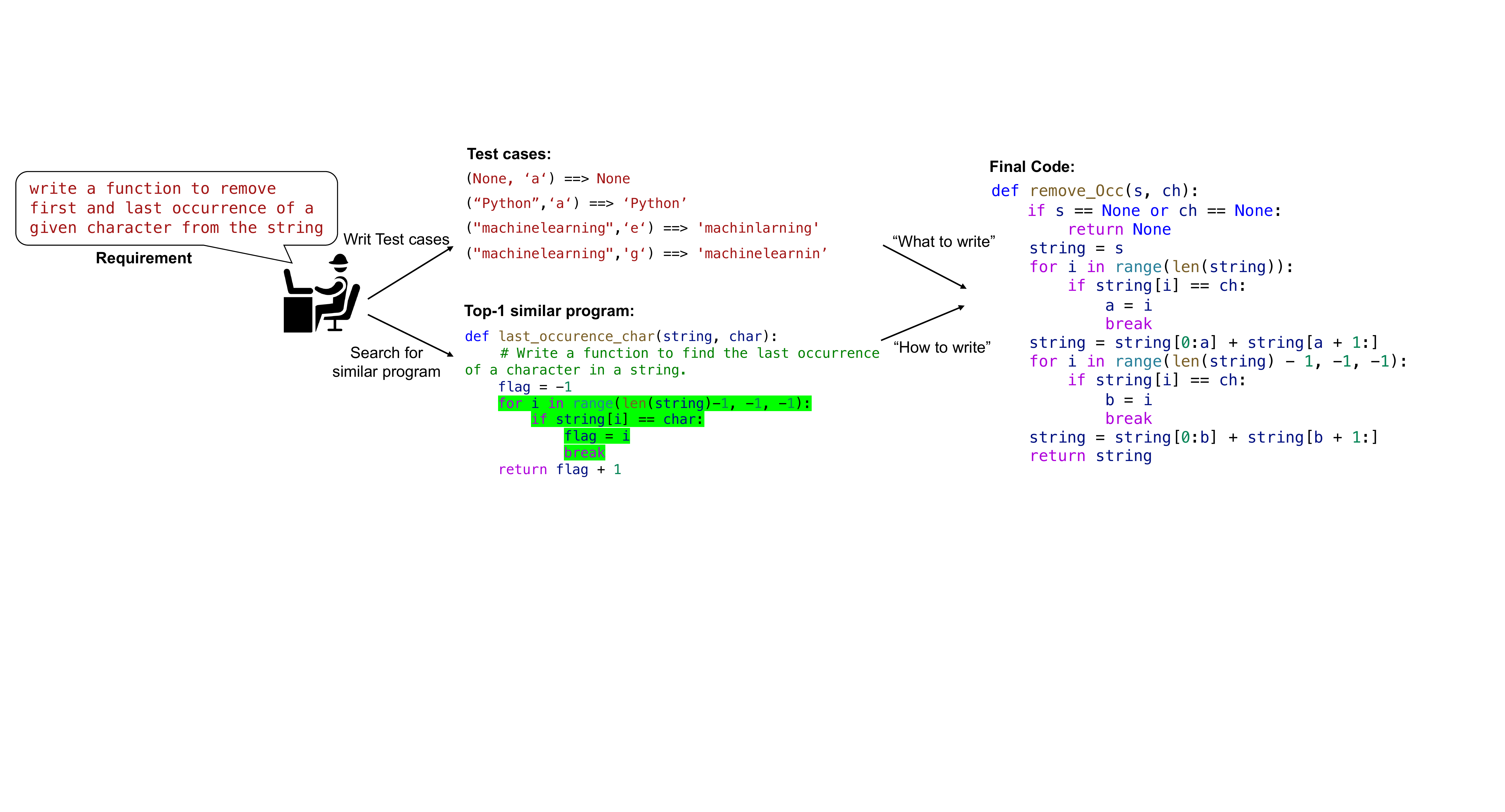}
\vspace{-0.2cm}
\caption{A motivating example of example retrieval.}
\label{fig:motivating_ex2}
\end{figure}

\textbf{Code Implementation $\rightarrow$ Example Retrieval.}
After understanding the input requirement, how to implement the code is challenging. It requires LLMs to use various algorithms or libraries. Figure \ref{fig:motivating_ex2} (a) and (b) show a requirement from a real-world benchmark \cite{MBPP} and its unit test for evaluation, respectively. We select Codex as the base model. Figure \ref{fig:motivating_ex2} (c) shows a program generated by few-shot prompting. The program contains a wrong condition statement highlighted in yellow. This is because the model does not know how to judge a string containing lowercase letters joined with an underscore.

To alleviate the above problem, we propose \textit{example retrieval}.
Our motivation is that human developers often search for similar programs and learn programming skills from them. Figure \ref{fig:motivating_ex2} (d) shows some retrieved programs based on the similarity of requirements. The retrieval metric is the BM25 score. We sort the results in descending order of BM25 score. 
We can see that the retrieved programs contain lots of relevant content (\eg \texttt{re.search}), which benefits code implementation. Thus, we design a retriever to search for similar programs as examples in prompts. We expect LLMs can learn from similar programs how to implement new programs.

Since the maximum input length of LLMs is usually limited (\eg 1024 tokens), the number of examples in a prompt is limited. Thus, we need to further select a set of programs from retrieved results as examples. 
A straightforward idea is to pick top similar programs as examples. However, as the programs are retrieved independently, we
find that retrieved results may contain redundant programs. 
In Figure \ref{fig:motivating_ex2} (d), Program-1, Program-2, and Program-3 are redundant, as all of them provide an API \texttt{re.search}, which teaches how to search a pattern in the text. Program-4 contains a relevant regular expression, which tells how to design a pattern. Suppose the number of examples is 2. The examples will contain redundant programs (\ie Program-1\&2) and miss more informative Program-4.

Thus, we design a selector for selecting examples, which can filter out redundant programs in retrieved results. Suppose the number of examples is 2. In Figure \ref{fig:motivating_ex2} (d), our selector will select Program-1 and Program-4 as examples. 
Figure \ref{fig:motivating_ex2} (e) shows a program generated by \method. It successfully learns how to write regular expressions from Program-4 and learns how to use \texttt{re.search} to find patterns from Program-1.

\begin{figure*}[t]
\centering
\includegraphics[width=\linewidth]{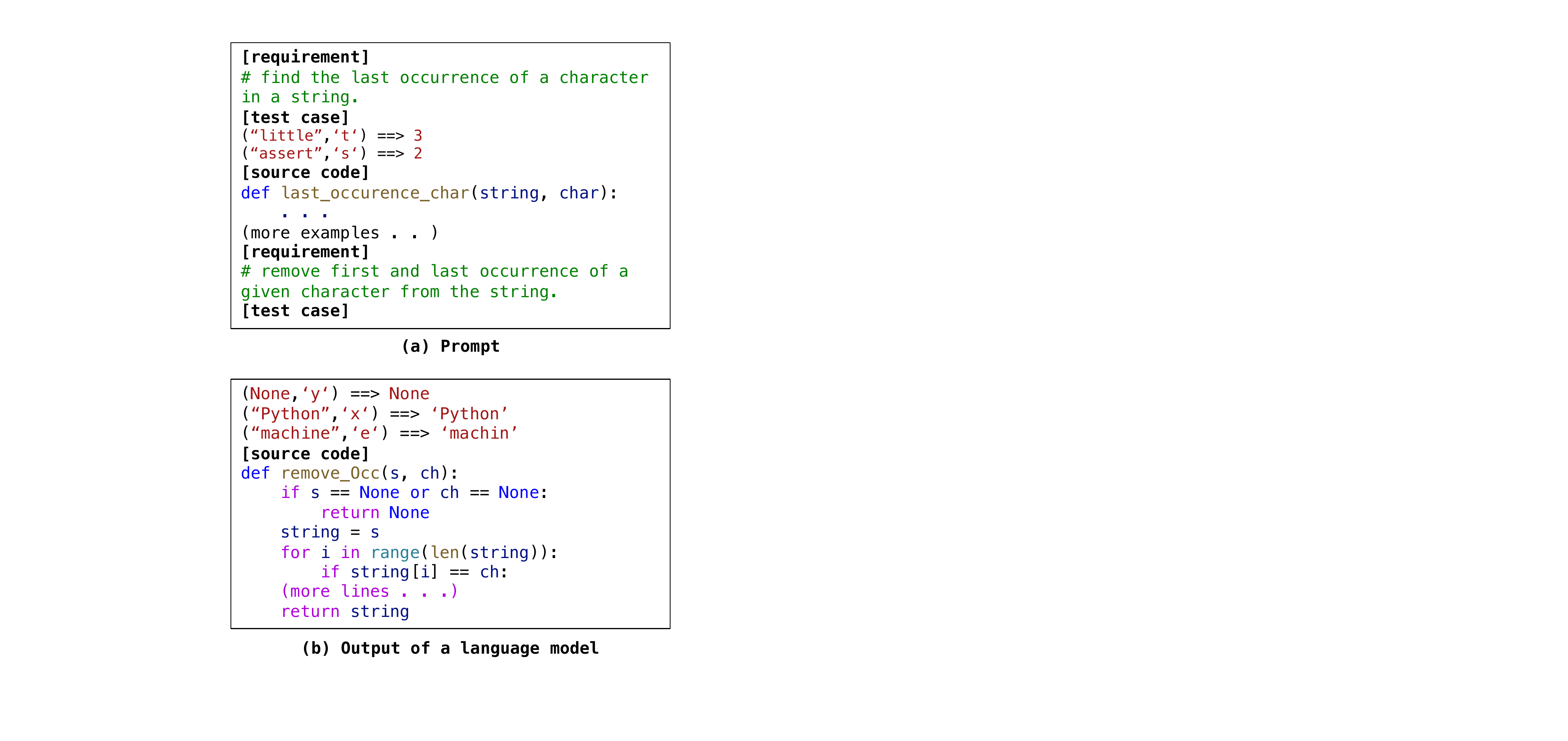}
\vspace{-0.4cm}
\caption{An overview of \method. Given a requirement, it selects examples from similar programs and constructs a prompt. LLMs first output an intermediate preliminary and then generate the source code. $x$, $y$, and $t$ denote requirements, programs, and intermediate preliminaries, respectively.}
\label{fig:overview}
\end{figure*}

\section{Methodology}
\label{sec:approach}

In this section, we propose a novel prompting technique for code generation, named \method. In the subsections, we first present an overview of \method and then describe its details.

\subsection{An Overview}
\label{sec:approach:overview}

Code generation aims to generate the source code $y$ based on a natural language requirement $x$. \method leverages large language models (LLMs) to generate programs via prompting. Figure \ref{fig:overview} shows an overview of \method during inference. Given an input requirement $x_{test}$, \method generates code in three steps.
\begin{itemize}[leftmargin=*]
    \item \textbf{Example Retrieval.} 
    It uses a \textit{retriever} and a \textit{selector} to select $k$ similar $<$requirement, code$>$ pairs ($\{x_i, y_i\}_{i=1}^k$) from a retrieval corpus as examples.
    \item \textbf{Prompt Construction.} 
    It employs an \textit{analyzer} to convert examples into $<$requirement, preliminary, code$>$ triples ($\{x_i, a_i, y_i\}_{i=1}^k$). A preliminary is a software artifact for clarifying the requirement, such as test cases. The examples are concatenated with the input requirement to construct a prompt.
    \item \textbf{Code Generation.} The prompt is fed into LLMs. By learning from examples, LLMs first output an intermediate preliminary and then generate the code.
\end{itemize}
where $x_i, y_i, a_i$ denote the requirement, the code, and the preliminary in $i$-th example, respectively.

\subsection{Example Retrieval}
\label{sec:approach:example_retrieval}

As shown in Figure \ref{fig:overview}, the first step has two goals: (i) retrieve similar programs and (ii) select a few examples from retrieved programs. We design a \textit{retriever} and a \textit{selector} to achieve these goals, respectively. The details of the two modules are shown as follows. 

\subsubsection{Retriever}

Similar programs often have similar natural language requirements \cite{Retrieve_edit_code_generation,SkCoder}. There, we take the input requirement as a query to search for similar requirements in a retrieval corpus. Then, we extract the corresponding programs as similar programs.

Specifically, we leverage an open-source search engine named Lucene \cite{Lucene} to build our retriever and use the training data as a retrieval corpus. We employ the BM25 score \cite{BM25} as the retrieval metric, which is widely used in previous studies \cite{Re2Com, EditSum}. The BM25 score is a bag-of-words retrieval function and is used to estimate the lexical-level similarity of two sentences. The more similar the two sentences are, the higher the value of BM25 scores.
In this paper, the retriever outputs top-$m$ similar programs based on the BM25 score.

The reason for choosing BM25+Lucene is that they can achieve good retrieval accuracy and have low complexity. Considering that the retrieval corpus is often large-scale, a lightweight retriever is closer to practical applications. 
In Section \ref{sec:results}-RQ5, we also explore other designs for the retriever and compare them to our design.

\begin{figure}[t]
\centering
\includegraphics[width=0.9\linewidth]{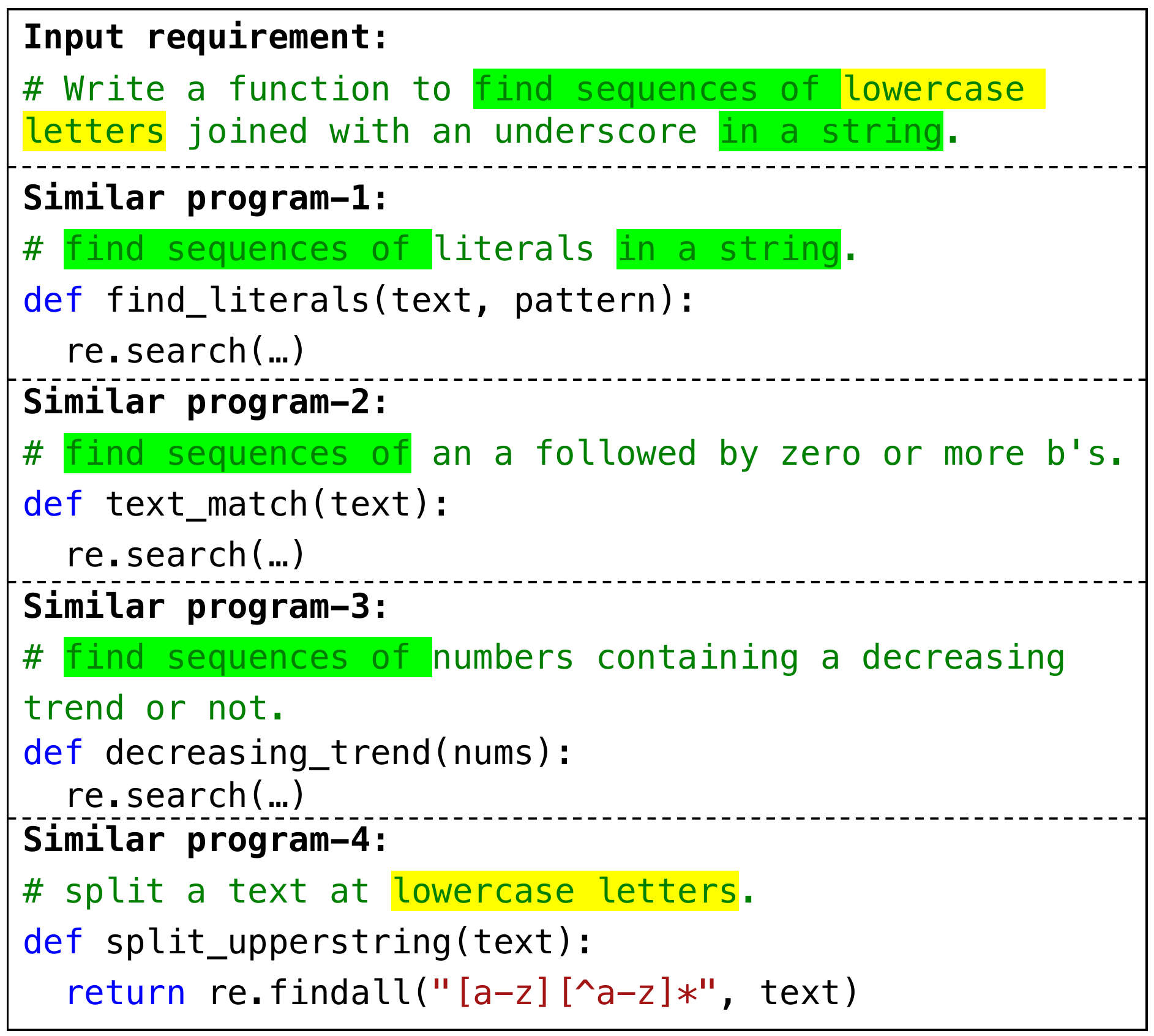}
\vspace{-0.3cm}
\caption{A requirement and its similar programs.}
\label{fig:selector_ex}
\end{figure}

\begin{algorithm}[t]
    \caption{The algorithm of our selector.}
    \label{algo:selector}
    \begin{algorithmic}[1]
        \REQUIRE~~\\
            Input requirement $x_{test}$, similar programs $\{(x_i,y_i)\}_{i=1}^m$; \\
            The number of examples $k, k<=m$, decay factor $\lambda$. \\
        \ENSURE~~\\
            Selected examples $T, \{(x_i,y_i)\}_{i=1}^k$.
        \STATE $T \gets \text{Empty Ordered List}$ 
        \STATE $S \gets {\rm Extract\_ngrams\_with\_count}(x_{test})$ \label{equ:extract_ngram}
        \FOR{$i$ in $\{1,\cdots,m\}$}
            \STATE $Q[i] \gets {\rm Extract\_ngrams\_with\_count}(x_i)$
        \ENDFOR
        \WHILE{$len(T)<k$}
            \FOR{$i$ in $\{1,\cdots,m\}$}
                \STATE $Score[i] \gets {\rm Ngram\_overlap\_score}(S,Q[i])$ \label{equ:ngram_score}
            \ENDFOR
            \STATE $j \gets {\rm argmax(Score)}$
            \STATE $T.append((x_j, y_j))$
            \STATE $matched\_ngrams \gets S \cap Q[j]$
            \STATE $Q[j] \gets \emptyset$
            \FOR{$ngram \in match\_ngrams$}
                \STATE $S[ngram] \times= \lambda$
            \ENDFOR
        \ENDWHILE
        \RETURN $T$
    \end{algorithmic}
\end{algorithm}

\subsubsection{Selector}
We can obtain top-$m$ similar programs from the retriever. However, the maximum input length of LLMs (\eg 1024 tokens) and the inference budget are often limited. It leads that the number of examples (\ie $k$) in a prompt is also limited (\eg three examples). It is necessary to further select $k$ programs from retrieved results as examples.

A straightforward idea is to pick top-$k$ similar programs as examples. However, as the programs are scored independently, we find that retrieved results may contain redundant programs.
Figure \ref{fig:selector_ex} shows a requirement and its similar programs. Similar programs are ranked by the BM25 score.
We can see that top-3 programs are redundant, as all of them use an API (\ie \texttt{re.search}) to find sequences of a specific pattern. 
Program-4 contains a relevant regex expression. However, as Program-4 has fewer overlapping $n$-grams with the input requirement, it has a relatively small BM25 score. 
Obviously, directly selecting top-$k$ (\eg top-3) retrieved programs is unreasonable, as it will introduce redundant programs and ignore more informative Program-4.

In this paper, we design a selector, which can filter out redundant programs in retrieved results. The algorithm of the selector is shown in Algorithm \ref{algo:selector}. We first extract all $n$-grams of the input requirement and all similar requirements (lines 2-5). In this paper, $n$ is set to 4 by default. Then, we calculate a recall-based ROUGE-$n$ score between the input requirement and each similar requirement using the following equations (lines 7-9).
\begin{align}
& R_{n} = \frac{\sum_{n\_gram \in S \cap Q} S(n\_gram)}{\sum_{n\_gram \in S} S(n\_gram)} \\
& Score = \exp(\frac{1}{n} \sum_n \log(R_{n}))
\end{align}
We get a similar requirement with the maximum score and add its corresponding program to examples (lines 10-11). Then, the matched $n$-grams between the similar requirement and the input requirement are decayed by a factor $\lambda$. This process (lines 6-17) is repeated until the number of examples reaches the upper bound.
The motivation for the decay is to filter out redundant programs, \ie programs with the same matched $n$-grams. For example, in Figure \ref{fig:selector_ex}, we first add Program-1 to examples and then decay its matched $n$-grams (\eg \texttt{find sequences of}). Subsequent programs with the same matched $n$-grams (\ie Program-2 and Program-3) are considered redundant and will be ignored. Program-4 contains new matched $n$-grams (\eg \texttt{lowercase letters}) and probably contains new information. Thus, Program-4 will obtain a higher score and is added to the examples.

By the above process, our selector filters out redundant programs and selects $k$ similar programs as examples. In practice, $m$ and $k$ are small numbers, such as $m=50,n=3$. Thus, the time complexity of our selector is acceptable. 

\begin{figure}[t]
\centering
\includegraphics[width=0.9\linewidth]{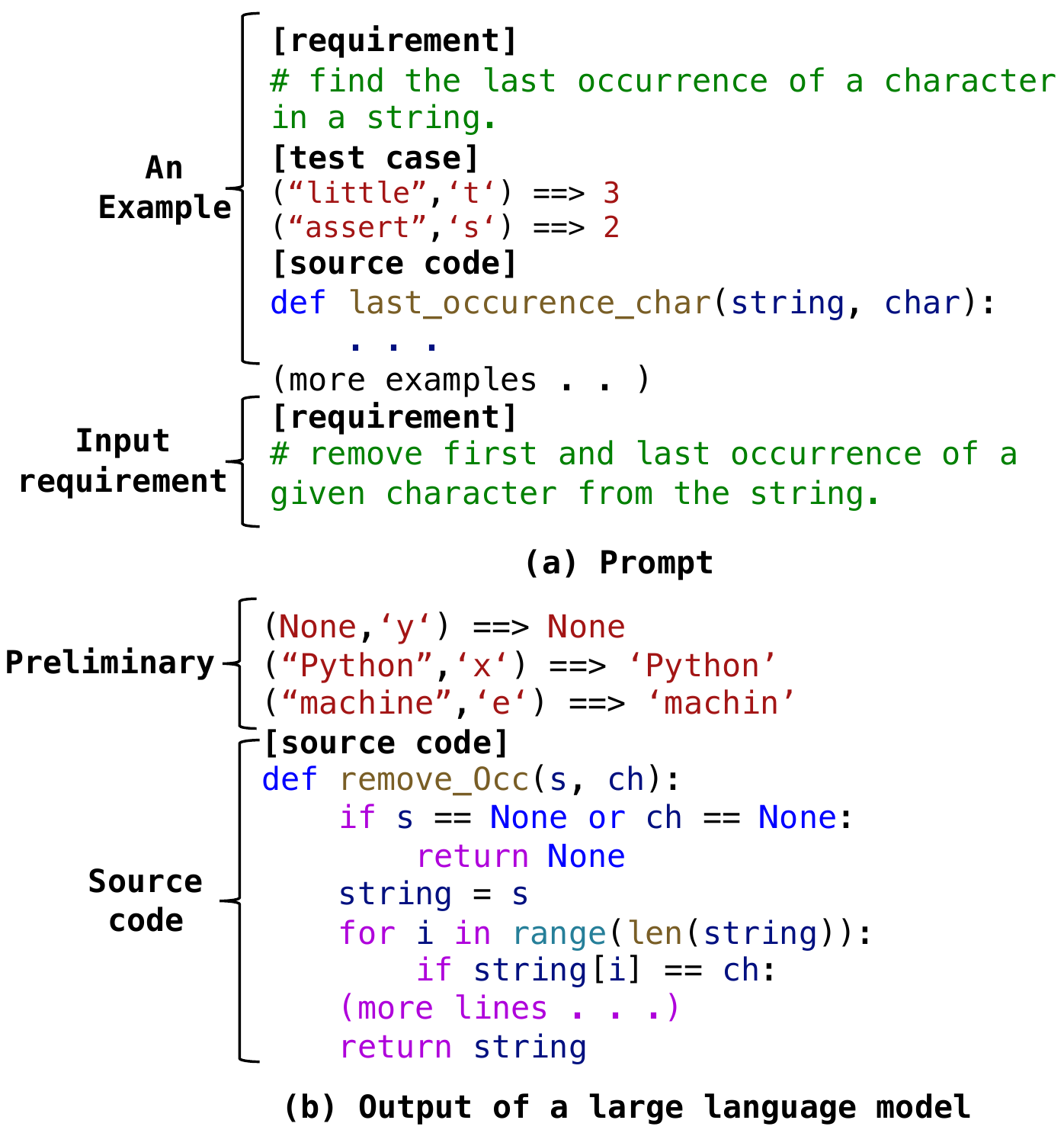}
\vspace{-0.2cm}
\caption{Examples of our prompt and an LLM's output.}
\label{fig:prompt}
\end{figure}

\subsection{Prompt Construction}
\label{sec:approach:prompt_construction}

The goal of this step is to construct a prompt. 
As stated in Section \ref{sec:introduction}, our guided code generation expects that LLMs can first output an intermediate preliminary and then generate the final code. To achieve this goal, we design a special prompt consisting of triple examples (\ie $<$requirement, preliminary, code$>$).

Specifically, we first use an analyzer to introduce preliminaries $\{t_i\}_{i=1}^k$ into selected examples $\{x_i,y_i\}_{i=1}^k$, obtaining triple examples $\{x_i,t_i,y_i\}_{i=1}^k$. The preliminary is a software artifact for clarifying requirements.
Inspired by test-driven development \cite{Test_Driven_Development}, this paper considers test cases as the preliminary by default. We also explore other choices (\eg APIs, method signature) in our experiments (Section \ref{sec:results}-RQ5). Then, we concatenate these triple examples with the input requirement to construct a prompt.

Figure \ref{fig:prompt} (a) shows an example of our prompt. The prompt begins with several examples and ends with a new requirement. \texttt{[requirement]}, \texttt{[test case]}, and \texttt{[source code]} are special tags that mark different parts in a triple.

We assume that test cases of examples are available. We think this assumption is acceptable. The reasons are two-fold. 
First, there are many public code generation datasets containing test cases, \eg MBPP \cite{MBPP} (474 samples), APPS \cite{APPS} (5,000 samples), and CodeContest \cite{AlphaCode} (13,328 samples). We can extract training data from these datasets and construct a retrieval corpus. 
Second, test-driven software development is popular in real-world scenarios. We can mine software repositories from open-source communities (\eg GitHub \cite{GitHub}) and extract code snippets equipped with test cases.

\subsection{Code Generation}
\label{sec:approach:code_generation}

In this step, we leverage an LLM to generate code based on the prompt. Following previous studies \cite{Codex, CodeGen, CodeGeeX, InCoder}, we view the LLM as a black-box generator and use it to complete the prompt. By learning from examples in the prompt, LLMs will first output a preliminary (\eg test cases) and then generate the code based on the preliminary and input requirement.

Figure \ref{fig:prompt} (b) shows an output of an LLM - CodeGeeX \cite{CodeGeeX}. We can see that CodeGeeX first generates some test cases and then implements a Python function. The test cases provide lots of valuable information (\eg input-output formats, invalid inputs) and guide the subsequent code generation.

\section{Study Design}
\label{sec:study_design}

To assess \method, we perform a large-scale study to answer six research questions. In this section, we describe the details of our study, including datasets, evaluation metrics, baselines, and base large language models (LLMs).

\subsection{Research Questions}
\label{sec:study:RQ}

Our study aims to answer the following research questions (RQs).

\textbf{RQ1: How does \method perform compared to existing prompting techniques?} 
This RQ aims to validate that \method has higher accuracy than existing prompting techniques in code generation. We apply \method and baselines to three LLMs and measure their accuracy on three code generation benchmarks. The evaluation metric is Pass@$K$.

\textbf{RQ2: How does \method perform compared to retrieval-based models?}
\method retrieves similar programs as examples in prompts. Some existing studies \cite{REDCODER,Jigsaw} also introduce information retrieval to augment code generation. In this RQ, we compare \method to these retrieval-based models. The evaluation metric is Pass@$K$.

\textbf{RQ3: Do human developers prefer code generated by \method?}
The ultimate goal of code generation is to assist human developers in writing code. In this RQ, we hire 10 developers (including industry employees and academic researchers) to manually review the code generated by \method and baselines. We measure the quality of code in three aspects, including correctness, code smell, and maintainability.

\textbf{RQ4: What are the contributions of different modules in \method?}
\method contains three modules, \ie a retriever, a selector, and an analyzer. This RQ is designed to analyze the contributions of three modules to the performance. We select a base model, gradually introduce three modules, and observe the fluctuations in accuracy.

\textbf{RQ5: What are the better designs for three modules?}
This RQ aims to validate the superiority of our designs for three modules in \method. Specifically, we explore multiple designs for three modules and compare them to our designs.

\begin{table}[t]
\centering
\caption{Statistics of the datasets in our experiments.}
\vspace{-0.3cm}
\begin{tabular}{lccc}
\toprule
Statistics                 & MBPP & MBJP & MBJSP  \\
\midrule
Language                   & Python & Java & JavaScript \\
\midrule
\# Train                   & 384  & 383   & 383  \\
\# Dev                     & 90   & 90    & 90   \\
\# Test                    & 500  & 493   & 493  \\
\midrule
Avg. tokens in requirement & 16.50       & 16.71  & 16.53 \\
Avg. tokens in code        & 92.68       & 247.79 & 100.75 \\
\bottomrule
\end{tabular}
\label{tab:dataset}
\end{table}

\subsection{Evaluation Datasets and Metrics}
\label{sec:study:dataset}

\subsubsection{Datasets}
We conduct experiments on three public code generation benchmarks, including the MBPP in Python, MBJP in Java, and MBJSP in JavaScript. The statistics of the datasets are shown in Table \ref{tab:dataset}. The details of the datasets are described as follows.

\begin{itemize}[leftmargin=*]
    \item \textbf{MBPP \cite{MBPP}} contains 974 real-world programming problems that are constructed by crowd-sourcing. Each problem contains a natural language requirement, a single Python function, and three test cases. 
    \item \textbf{MBJP \cite{MBXP}} and \textbf{MBJSP \cite{MBXP}} both contain 966 crowd-sourced programming problems in Java and JavaScript, respectively. Each problem consists of a natural language requirement, an individual function, and 3 test cases.
\end{itemize}

\subsubsection{Metrics}
 
Following previous code generation studies \cite{Codex,CodeGeeX,CodeGen,InCoder}, we employ Pass@$k$ as our evaluation metric. Specifically, we generate $k$ programs for each requirement. A requirement is considered solved if any generated programs pass all test cases. We compute the percentage of solved requirements in total requirements as Pass@$k$. In this paper, $k$ is set to 1, 3, and 5.

We notice that previous studies \cite{Retrieve_edit_code_generation, CodeT5} also use some match-based metrics (\eg BLEU \cite{BLEU}). These metrics are initially designed for natural language generation and are poor in measuring the functionality of programs \cite{Codex}. Thus, we omit them in experiments.

\begin{table*}[t]
\centering
\caption{The results of \method and prompting baselines on three datasets. The values in parentheses are the relative improvements compared to the SOTA baseline - few-shot prompting.}
\vspace{-0.3cm}
\resizebox{\linewidth}{!}{
\begin{tabular}{ccccccccccc}
\toprule
\multirow{2}{*}{Base model} & \multirow{2}{*}{Prompting Technique}  & \multicolumn{3}{c}{MBPP}                                                  & \multicolumn{3}{c}{MBJP}   & \multicolumn{3}{c}{MBJSP} \\
                        &      & \multicolumn{1}{c}{Pass@1} & \multicolumn{1}{c}{Pass@3} & \multicolumn{1}{c}{Pass@5} & \multicolumn{1}{c}{Pass@1} & \multicolumn{1}{c}{Pass@3} & \multicolumn{1}{c}{Pass@5} & \multicolumn{1}{c}{Pass@1} & \multicolumn{1}{c}{Pass@3} & \multicolumn{1}{c}{Pass@5} \\ \midrule
\multirow{4}{*}{CodeGeeX-13B}    & Zero-shot prompting  & 5.20  & 13.80 & 19.40 & 4.46  & 11.97 & 18.26  & 0.20  & 0.20 & 0.41   \\
                        & CoT prompting & 12.60 & 23.40 & 30.20 & 14.40 & 28.19 & 33.67 & 11.35 & 21.10 & 25.96 \\
                        & Few-shot prompting  & 20.40  & 30.60 & 36.00 & 16.63  & 26.17 & 34.48  & 11.16  & 19.88 & 25.56  \\
                        & \method   & \textbf{26.74 ($\uparrow$ 31.1\%)}  & \textbf{36.43 ($\uparrow$ 19\%)} & \textbf{41.13 ($\uparrow$ 14.2\%)} & \textbf{28.38 ($\uparrow$ 70.7\%)}  & \textbf{36.79 ($\uparrow$ 40.6\%)} & \textbf{41.54 ($\uparrow$ 20.5\%)} & \textbf{21.03 ($\uparrow$ 88.4\%)} & \textbf{31.44 ($\uparrow$ 58.2\%)} & \textbf{36.04 ($\uparrow$ 41\%)} \\ \midrule
\multirow{4}{*}{CodeGen-6B}    & Zero-shot prompting & 10.40  & 19.40 & 24.40 & 14.81 & 25.76 & 31.44 & 8.72 & 19.67 & 22.92   \\
                        & CoT prompting & 13.00 & 21.00 & 26.00 & 13.59 & 25.35 & 31.24 & 11.56 & 20.08 & 24.54 \\
                        & Few-shot prompting  & 14.60  & 24.00 & 30.20 & 18.25 & 30.02 & 34.68 & 9.94 & 19.88 & 23.12  \\
                        & \method   & \textbf{22.83 ($\uparrow$ 56.4\%)}  & \textbf{34.58 ($\uparrow$ 44.1\%)} & \textbf{40.16 ($\uparrow$ 33\%)} & \textbf{22.45 ($\uparrow$ 23\%)} & \textbf{34.27 ($\uparrow$ 14.2\%)} & \textbf{40.96 ($\uparrow$ 18.1\%)} & \textbf{16.45 ($\uparrow$ 65.5\%)} & \textbf{27.31 ($\uparrow$ 37.4\%)} & \textbf{32.16 ($\uparrow$ 39.1\%)}  \\ \midrule
\multirow{4}{*}{InCoder-6B}    & Zero-shot prompting & 4.20  & 11.40 & 16.20 & 2.23  & 5.88 & 9.13 & 3.65  & 5.88 & 8.11 \\
                        & CoT prompting & 3.99 & 10.65 & 15.31 & 1.83 & 4.46 & 7.10 & 1.22 & 2.03 & 4.67 \\
                        & Few-shot prompting  & 12.80  & 22.80 & 28.20 & 10.95  & 23.53 & 26.17  & 12.78  & 22.52 & 27.79 \\
                        & \method     & \textbf{20.16 ($\uparrow$ 57.5\%)}  & \textbf{31.44 ($\uparrow$ 37.9\%)} & \textbf{34.10 ($\uparrow$ 20.92\%)} & \textbf{16.37 ($\uparrow$ 49.5\%)} & \textbf{29.89 ($\uparrow$ 27\%)} & \textbf{34.74 ($\uparrow$ 32.7\%)}  & \textbf{15.97 ($\uparrow$ 25\%)}  & \textbf{27.13 ($\uparrow$ 20.5\%)} & \textbf{30.65 ($\uparrow$ 10.3\%)} \\
\bottomrule
\end{tabular}}
\label{tab:RQ1}
\vspace{-0.3cm}
\end{table*}

\subsection{Comparison Baselines}
\label{sec:study:baseline}

This paper is to propose a new prompting technique for code generation. Thus, we select three existing prompting techniques as baselines.
\begin{itemize}[leftmargin=*]
    \item \textbf{Zero-shot prompting} \cite{CodeGen,Codex} directly feeds the input requirement into LLMs. Then, it extracts the code from LLMs' outputs.
    \item \textbf{Few-shot prompting} \cite{Codex} randomly selects several $<$requirement, code$>$ pairs as examples and constructs a prompt, which is fed into an LLM. Then, it extracts the code from LLMs' outputs.
    \item \textbf{Chain-of-Thought (CoT) prompting} \cite{Chain-of-thought} is a variant of few-shot prompting. CoT prompting asks LLMs first to generate a series of intermediate natural language reasoning steps and then output the code.
\end{itemize}

\method retrieves similar programs to assist LLMs in generating code. Some studies also introduce information retrieval to augment code generation. We compare \method to these retrieval-based models.
\begin{itemize}[leftmargin=*]
    \item \textbf{REDCODER} \cite{REDCODER} retrieves similar programs and fine-tunes a pre-trained model - PLBART \cite{PLBART} to generate code based on the requirement and similar programs. 
    \item \textbf{Jigsaw} \cite{Jigsaw} searches for similar programs from API documentation and insert them into the prompts.
\end{itemize}

\subsection{Base Large Language Models}
\label{sec:study:LLMs}

We select three open-source LLMs as base models. The details of the base models are shown as follows.

\begin{itemize}[leftmargin=*]
    \item \textbf{CodeGeeX \cite{CodeGeeX}} is a multilingual LLM for source code with 13 billion parameters. CodeGeeX is pre-trained with a large corpus of more than 20 programming languages (\eg  Python, Java, JavaScript). We download the model weight from the official website \cite{CodeGeeX-web} and run CodeGeeX according to official instructions.
    \item \textbf{CodeGen \cite{CodeGen}} is a family of LLMs for source code that is pre-trained with extensive natural language and code data. We select CodeGen-Multi-6.1B (CodeGen-6B) as a base model.
    \item \textbf{InCoder \cite{InCoder}} is a multilingual LLM for code generation. It is pre-trained with 216 GB of code data. We use a version with 6.7 billion parameters (InCoder-6B) as a base model.
\end{itemize}

The reason why we do not choose the GPT series of models (\eg ChatGPT \cite{ChatGPT}) as the base models is that they are closed source. Although we can access GPT models through the OpenAI's APIs, these models are likely to be updated dynamically, affecting the fairness and reproducibility of experiments. Thus, we leave them to future work.

\subsection{Implementation Details}
\label{sec:study:implementation}

\textbf{Example Retrieval.} For each dataset, the retrieval corpus is its training data. We exclude the ground truths from the outputs of our retriever. We first retrieve top-20 similar programs and then use the selector to select three examples. For ensuring fairness, the number of examples in \method and baselines is the same. 

\textbf{Prompt Construction.} In experimental datasets, the retrieval corpus (\ie training data) has been equipped with test cases by data collector \cite{MBPP, MBXP}. Thus, the analyzer utilizes pre-defined rules to extract test cases and transform retrieved programs into <requirement, test cases, code> triples.

\textbf{Code Generation.} Following previous studies \cite{Codex, CodeGen, InCoder}, we use nucleus sampling \cite{nucleus_sampling} to decode programs from LLMs. The temperature is 0.8 and the top-$p$ is 0.95. The maximum generated lengths are 400, 500, and 500, respectively. The sampling settings of baselines are the same as the ones of \method.

\section{Results and Analyses}
\label{sec:results}

In the first research question, we evaluate the performance of \method with respect to existing prompting techniques.

\begin{table*}[t]
\centering
\caption{The comparison of retrieval-based baselines and \method. The values in parentheses are relative improvements compared to the SOTA baseline - Jigsaw.}
\vspace{-0.3cm}
\resizebox{\linewidth}{!}{
\begin{tabular}{ccccccccccc}
\toprule
\multirow{2}{*}{Approach}  & \multicolumn{3}{c}{MBPP}                                                  & \multicolumn{3}{c}{MBJP}   & \multicolumn{3}{c}{MBJSP}  \\
                        & \multicolumn{1}{c}{Pass@1} & \multicolumn{1}{c}{Pass@3} & \multicolumn{1}{c}{Pass@5} & \multicolumn{1}{c}{Pass@1} & \multicolumn{1}{c}{Pass@3} & \multicolumn{1}{c}{Pass@5} & \multicolumn{1}{c}{Pass@1} & \multicolumn{1}{c}{Pass@3} & \multicolumn{1}{c}{Pass@5} \\ \midrule
REDCODER & 3.37 & 6.21 & 9.74 & 4.46  & 7.51 & 9.94 & 4.87 & 10.34 & 12.78  \\
Jigsaw  & 23.65  & 33.97 & 37.78  & 22.99 & 33.26 & 36.95 & 18.16 & 28.79 & 34.08 \\
\method & \textbf{26.74 ($\uparrow$ 13.1\%)}  & \textbf{36.43 ($\uparrow$ 7.2\%)}   & \textbf{41.13 ($\uparrow$ 8.9\%)}    & \textbf{28.38 ($\uparrow$ 23.44\%)}   & \textbf{36.79 ($\uparrow$ 10.61\%)}  & \textbf{41.54 ($\uparrow$ 12.42\%)} & \textbf{21.03 ($\uparrow$ 15.8\%)} & \textbf{31.44 ($\uparrow$ 9.2\%)} & \textbf{36.04 ($\uparrow$ 5.8\%)}  \\ 
\bottomrule
\end{tabular}}
\label{tab:RQ2}
\end{table*}

\noindent\textbf{RQ1: How does \method perform compared to existing prompting techniques?} 

\textbf{Setup.} We apply \method and three prompting baselines to three base models (Section \ref{sec:study:LLMs}). Then, we use Pass@k to measure their performance on three benchmarks (Section \ref{sec:study:dataset}).

\textbf{Results.} The results on three benchmarks are shown in Table \ref{tab:RQ1}. The values in parentheses are relative improvements compared to the SOTA baseline - few-shot prompting.

\textbf{Analyses.}
\uline{(1) \method performs better than baselines on three benchmarks.}
Compared to the SOTA baseline - few-shot prompting, in terms of Pass@1, \method outperforms it by up to 56.4\% in MBPP, 70.7\% in MBJP, and 88.4\% in MBJSP. Pass@1 is a very strict metric and it is difficult to improve.
The significant improvements prove the superiority of \method in
code generation.
We attribute the improvements to our novel techniques, \ie example retrieval and guided code generation. The retrieved examples contain many relevant code elements teaching LLMs ``how to write''. Guided code generation asks LLMs to analyze requirements that tell LLMs ``what to write''. 
\uline{(2) \method is effective in LLMs with different sizes and different programming languages.}
Compared to few-shot prompting, in terms of Pass@1, \method improves CodeGeeX-13B by up to 88.4\%, CodeGen-6B by up to 65.5\%, and InCoder-6B by up to 57.5\%. In particular, we find that an LLM with \method even outperforms larger LLMs. For example, in the MBJSP, InCoder-6B with \method outperforms CodeGeeX-13B with few-shot prompting. It proves the potential of \method.
Besides, \method is language-agnostic and is effective in multilingual code generation (\ie Python, Java, and JavaScript).

\begin{tcolorbox}[size=title]
\textbf{Answer to RQ1:} \method outperforms existing prompting techniques on three benchmarks. In terms of Pass@1, \method outperforms the SOTA baseline by up to 56.4\% in MBPP, 70.7\% in MBJP, and 88.4\% in MBJSP. Besides, \method is effective in LLMs with different sizes. It improves CodeGeeX-13B by up to 88.4\%, CodeGen-6B by up to 65.5\%, and InCoder-6B by up to 57.5\%. The significant improvements prove the effectiveness of \method in code generation.
\end{tcolorbox}

\textbf{RQ2: How does \method perform compared to retrieval-based models?}

\textbf{Setup.}
In this RQ, we compare \method to two retrieval-based baselines, including REDCODER \cite{REDCODER} and Jigsaw \cite{Jigsaw}. Baselines and \method use the same retrieval corpus. Because REDCODER requires fine-tuning, we follow the official instructions and use the training data to train REDCODER.

\textbf{Results.}
The results on three benchmarks are shown in Table \ref{tab:RQ2}. The values in parentheses are relative improvements compared to
the SOTA baseline - Jiagsaw.

\textbf{Analyses.}
\uline{\method outperforms retrieval-based baselines in three benchmarks.} Compared to the SOTA baseline - Jigsaw, in terms of Pass@1, \method outperforms it by up to 13.1\% in MBPP, 23.44\% in MBJP, and 15.8\% in MBJSP. Jigsaw also retrieves similar programs for making prompts. The improvements show the effectiveness of our selector and analyzer. The selector filters out redundant similar programs and further improves the quality of examples. The analyzer constraints LLMs to first analyze requirements and then generate code. Besides, we notice that REDCODER has poor accuracy in three benchmarks. This is because the training data is limited, and fine-tuning easily leads to overfitting. It validates our motivation that introducing similar programs by prompting is a more suitable approach to LLMs.

\vspace{-0.2cm}
\begin{tcolorbox}[size=title]
\textbf{Answer to RQ2:} \method outperforms retrieval-based baselines. Specifically, it outperforms the SOTA baseline - Jigsaw by up to 13.1\% in MBPP, 23.44\% in MBJP, and 15.8\% in MBJSP.
\end{tcolorbox}

\begin{table*}[t]
\centering
\caption{The results of ablation study. The values in parentheses are relative improvements compared to few-shot promopting.}
\vspace{-0.3cm}
\resizebox{\linewidth}{!}{
\begin{tabular}{cccccccccccc}
\toprule
\multirow{2}{*}{Retriever} & \multirow{2}{*}{Selector} & \multirow{2}{*}{Analyzer} & \multicolumn{3}{c}{MBPP}                                                  & \multicolumn{3}{c}{MBJP}   & \multicolumn{3}{c}{MBJSP}  \\
                        &                            &                           & \multicolumn{1}{c}{Pass@1 (\%)} & \multicolumn{1}{c}{Pass@3} & \multicolumn{1}{c}{Pass@5} & \multicolumn{1}{c}{Pass@1} & \multicolumn{1}{c}{Pass@3} & \multicolumn{1}{c}{Pass@5} & \multicolumn{1}{c}{Pass@1} & \multicolumn{1}{c}{Pass@3} & \multicolumn{1}{c}{Pass@5} \\ \midrule
    \usym{2715}   & \usym{2715}  & \usym{2715}  & 20.40   & 30.60   & 36.00   & 16.63  & 26.17   & 34.48 & 11.16 & 19.88 & 25.56    \\
    \usym{1F5F8} & \usym{2715}  & \usym{2715}   & 24.00 ($\uparrow$ 17.6\%)    & 34.60 ($\uparrow$ 13.1\%)  & 38.20 ($\uparrow$ 6.1\%)    & 23.35 ($\uparrow$ 40.4\%)   & 33.67 ($\uparrow$ 28.7\%)  & 37.22 ($\uparrow$ 7.9\%) & 18.66 ($\uparrow$ 67.2\%) & 29.18 ($\uparrow$ 46.8\%) & 34.89 ($\uparrow$ 36.5\%)   \\
    \usym{1F5F8}   & \usym{1F5F8}  & \usym{2715} & \textbf{24.89 ($\uparrow$ 22\%)}  & \textbf{35.02 ($\uparrow$ 14.4\%)}   & \textbf{39.14 ($\uparrow$ 8.7\%)}    & \textbf{25.03 ($\uparrow$ 50.5\%)}   & \textbf{34.47 ($\uparrow$ 31.7\%)}  & \textbf{39.24 ($\uparrow$ 13.8\%)} & \textbf{19.73 ($\uparrow$ 76.8\%)} & \textbf{30.16 ($\uparrow$ 51.7\%)} & \textbf{35.34 ($\uparrow$ 38.3\%)}   \\ 
    \usym{1F5F8}   & \usym{1F5F8}  & \usym{1F5F8} & \textbf{26.74 ($\uparrow$ 31.1\%)}  & \textbf{36.43 ($\uparrow$ 19\%)} & \textbf{41.13 ($\uparrow$ 14.2\%)} & \textbf{28.38 ($\uparrow$ 70.7\%)}  & \textbf{36.79 ($\uparrow$ 40.6\%)} & \textbf{41.54 ($\uparrow$ 20.5\%)} & \textbf{21.03 ($\uparrow$ 88.4\%)} & \textbf{31.44 ($\uparrow$ 58.2\%)} & \textbf{36.04 ($\uparrow$ 41\%)}   \\ 
\bottomrule
\end{tabular}}
\label{tab:ablation_study}
\vspace{-0.3cm}
\end{table*}

\noindent\textbf{RQ3: Do human developers prefer code generated by \method?} 

\textbf{Setup.} The ultimate goal of code generation is to assist human developers in writing code. Thus, we conduct a human evaluation to measure programs generated by \method and baselines.
We follow settings of human evaluation in previous studies \cite{AixBench, SkCoder}. We have carefully checked the evaluation settings and think our settings are reliable. We manually evaluate programs in three aspects:
\begin{itemize}[leftmargin=*]
    \item \textbf{Correctness (whether the program satisfies the given requirement).} 0 point: the program is totally inconsistent with the requirement. 1 point: the program is implemented, but misses some details. 2 points: the program is correctly implemented.
    \item \textbf{Code Smell (whether the program contains bad code smell).} 0 point: There are better solutions in terms of performance. Or there is serious code smell. 1 point: Some details are not in place. There is code smell of low severity. 2 points: No obviously better code in terms of performance exists. If possible, resources are released accordingly. No obvious code smell.
    \item \textbf{Maintainability (whether the implementation is standardized and has good readability).} 0 point: The program does not follow a consistent speciﬁcation, or there are many meaningless names in variable naming, or there are certain repetitions and redundant codes. 1 point: The program implementation meets certain specifications. But some variable names can be further refined. 2 points: The program implementation is relatively standardized, the variable naming is basically semantically straightforward, and the readability is better.
\end{itemize}

\begin{table}[t]
\caption{The results of human evaluation. The values in parentheses are the relative improvements compared to the SOTA baseline - few-shot prompting.}
\vspace{-0.3cm}
\resizebox{\linewidth}{!}{
\begin{tabular}{lccc}
\toprule
Approach        & Correctness & Code smell & Maintainability \\ \midrule
Zero-shot prompting     & 0.3167      & 1.1033       & 1.2749          \\
CoT prompting          & 0.6671      & 1.1405       & 1.4479          \\
Few-shot prompting     & 0.9769      & 1.2148       & 1.5420          \\
\method         & \textbf{1.5802 ($\uparrow$ 61.8\%)}      & \textbf{1.6241 ($\uparrow$ 33.7\%)}      & \textbf{1.7544 ($\uparrow$ 13.8\%)}         \\ \bottomrule
\end{tabular}}
\label{tab:human_evaluation}
\vspace{-0.2cm}
\end{table}

We explain the above aspects to evaluators through some examples.
After discussing with evaluators, we set the score of each aspect to an integer, ranging from 0 to 2 (from bad to good). 
For \method and baselines, we select a fixed base model (\ie CodeGen-2B) and collect 200 generated programs per approach. Finally, we obtain 1,000 programs for evaluation. 
We invite 10 developers with 3-5 years of development experience to evaluate the generated programs in the form of a questionnaire. The 1,000 code snippets are divided into 5 groups, with each questionnaire containing one group. 
The programs are randomly shuffled and anonymously reviewed by evaluators.
Each group is evaluated by two evaluators, and the final score is the average of two evaluators' scores. Evaluators are allowed to search the Internet for unfamiliar concepts.

\textbf{Results.} 
The results of the human evaluation are shown in Table \ref{tab:human_evaluation}.
The values in parentheses are the relative improvements compared to the SOTA baseline - few-shot prompting.

\textbf{Analyses.}
\uline{\method is better than all baselines in three aspects.} Specifically, our \method outperforms the SOTA baseline - few-shot prompting by 61.8\% in correctness, 33.7\% in code smell, and 13.8\% in maintainability. The improvements show that \method has better usability and is promising in practical applications.
Besides, all the p-values are substantially smaller than 0.05, which shows the improvements are statistically significant.

\begin{tcolorbox}[size=title]
\textbf{Answer to RQ3:} Human evaluation shows that human developers prefer programs generated by \method. It outperforms the SOTA baseline by 61.8\% in correctness, 33.7\% in code smell, and 13.8\% in maintainability.
\end{tcolorbox}

\vspace{-0.2cm}
\noindent\textbf{RQ4: What are the contributions of different modules in \method?}

\textbf{Setup.} 
\method contains three modules, \ie a retriever, a selector, and an analyzer. This RQ is designed to analyze the contributions of three modules to the performance.
We select CodeGeeX as the base model and conduct an ablation study by gradually adding three modules.

\textbf{Results.} The results are shown in Table 5. 
$\usym{1F5F8}$ and $\usym{2715}$ represent adding and removing corresponding modules, respectively.
Without three modules, the base model uses few-shot prompting to generate code. 
After adding a retriever, the base model selects top-$k$ similar programs as examples and directly generates code. 
After adding a selector, the base model selects $k$ examples from similar programs and then generates code.
After further introducing an analyzer, the base model uses \method to generate code.

\textbf{Analyses.}
\uline{All modules are necessary for \method to perform the best.}
After adding a retriever, the performance of the base models is improved. In terms of Pass@1, the retriever brings a 17.6\% improvement in MBPP, a 40.4\% improvement in MBJP, and a 67.2\% improvement in MBJSP. It validates our motivation that retrieved programs contain lots of useful information that benefits code generation. 
After adding a selector, the performance of the base model is further improved. It shows that our selector can effectively filter out redundant programs in retrieved results and improve the quality of examples. 
After further introducing an analyzer, the base model achieves better results. In terms of Pass@1, the base model is improved by 31.1\% in MBPP, 70.7\% in MBJP, and 88.4\% in MBJSP. It proves the effectiveness of guided code generation in analyzing requirements. 

\begin{tcolorbox}[size=title]
\textbf{Answer to RQ4:} Three modules are essential for the performance of \method. The performance of CodeGeeX on three benchmarks is substantially improvement by gradually adding three modules.
\end{tcolorbox}

\begin{table*}[t]
\caption{The performance of \method with different designs. ``w/'' is the abbreviation of with.}
\label{tab:RQ5}
\vspace{-0.3cm}
\begin{tabular}{lccccccccc}
\toprule
\multicolumn{1}{c}{\multirow{2}{*}{Approach}} & \multicolumn{3}{c}{MBPP} & \multicolumn{3}{c}{MBJP} & \multicolumn{3}{c}{MBJSP} \\
\multicolumn{1}{c}{} & Pass@1 & Pass@3 & Pass@5 & Pass@1 & Pass@3 & Pass@5 & Pass@1 & Pass@3 & Pass@5 \\ \midrule
\method & \textbf{26.74} & \textbf{36.43} & \textbf{41.13} & \textbf{28.38} & \textbf{36.79} & \textbf{41.54} & \textbf{21.03} & \textbf{31.44} & \textbf{36.04} \\
\quad w/ Dense retriever & 26.63 & 36.42 & 41.10 & 28.16 & 36.55 & 41.32 & 20.88 & 31.27 & 35.94 \\
\quad w/ BLEU selector & 25.61 & 35.71 & 40.74 & 27.86 & 35.91 & 40.77 & 20.15 & 30.42 & 35.47 \\
\quad w/ API analyzer & 25.10 & 35.24 & 40.38 & 26.44 & 35.16 & 40.12 & 19.86 & 30.23 & 35.41 \\
\quad w/ signature analyzer & 26.14 & 35.96 & 40.89 & 27.35 & 36.11 & 40.98 & 20.58 & 30.89 & 35.86 \\
\bottomrule
\end{tabular}
\vspace{-0.3cm}
\end{table*}

\vspace{-0.2cm}
\noindent\textbf{RQ5: What are the better designs for three modules in \method?}

\textbf{Setup.} 
As stated in Section \ref{sec:approach:overview}, \method contains three modules, \ie a retriever, a selector, and an analyzer. In this RQ, we explore different designs for three modules and validate the superiority of our designs. 
We select CodeGeeX as the base model.
The evaluation settings are shown as follows.

(1) A retriever takes the input requirement as a query and searches for similar programs from a retrieval corpus. We design two choices for the retriever:
\begin{itemize}[leftmargin=*]
    \item Dense retriever. It uses a neural encoder to convert the requirements into vector representations. Then, it retrieves similar programs based on the similarity of vector representations. In experiments, we use an off-the-shelf natural language representation model \cite{SentenceBERT} as the encoder.
    \item Sparse retriever (\method). As stated in Section \ref{sec:approach:example_retrieval}, it uses the BM25 score as the retrieval metric. BM25 score can measure the lexical-level similarity of two requirements.
\end{itemize}

(2) A selector aims to score similar programs and filter redundant programs. For the score function in the selector (line 8 of Algorithm \ref{algo:selector}), we design two choices:
\begin{itemize}[leftmargin=*]
    \item BLEU \cite{BLEU}. It extracts overlapping $n$-grams between the input requirement and the similar requirement. Then, it computes the precision of $n$-grams in the similar requirement.
    \item ROUGE-N \cite{ROUGE} (\method). It extracts overlapping $n$-grams between the input requirement and the similar requirement. Then, it computes the recall of $n$-grams in the input requirement.
\end{itemize}

(3) An analyzer is to introduce preliminaries into examples. A preliminary is a special software artifact that benefits the requirement understanding. For the preliminary, we design three choices:
\begin{itemize}[leftmargin=*]
    \item API sequence. APIs are important elements in code and reflect the functionality of the code. Pre-designing APIs help LLMs to think about how to solve requirements. We use a program analysis tool \cite{tree-sitter} to extract APIs from examples and view the API sequence as a preliminary (\eg \texttt{open, numpy.array, write}).
    \item Method signature. It contains input-output parameters and their types, which clearly indicate the inputs and outputs of requirements. Thus, we consider the method signature as a preliminary (\eg \texttt{def floor\_Min(A: int, B: int, N: int)}).
    \item Test cases (\method). Test cases exactly define the requirement, including the input-output format, edge cases, and functionality. We consider several test cases as the preliminary, such as \texttt{(``Python'',``o'') -> 1); (``little'',``t'') -> 2);}.
\end{itemize}

\textbf{Results and Analyses.} The results are shown in Table \ref{tab:RQ5}. ``w/'' is the abbreviation of with.
\uline{(1) A dense retriever is comparable to our retriever, but has a lower efficiency.} In Table \ref{tab:RQ5}, compared to \method, \method with dense retriever has a slight drop in performances. It indicates that code generation prefers lexically similar programs, which contain lots of reusable content. 
Similar findings can be found in code completion work \cite{ReAcc}.
Besides, the dense retriever has a higher complexity and is hard to be applied to a large-scale retrieval corpus.
\uline{(2) The BLEU selector prefers shorter examples and is suboptimal.} Compared to \method, \method with BLEU selector has an obvious decrease in accuracy. We inspect some failed samples and find that the BLEU selector prefers shorter examples. This is because BLEU is the precision of $n$-gram in similar requirements. The shorter the similar requirement, the higher the BLEU. It leads that the selector tends to select short programs as examples and ignores some informative but long examples.
\uline{(3) Test cases are more suitable to the preliminary than APIs and method signatures.} We carefully inspect some cases.  First, many requirements in benchmarks do not require APIs or only involve a few trivial APIs (\eg range, split, and len). It causes that generated APIs bring limited benefits to code generation. Second, by generating method signatures, LLMs are asked to think about the input-output format, which benefits code generation. But method signatures miss other necessary details, such as edge cases. \method considers test cases as the preliminary. Test cases are common in code files. Thus, it is feasible for LLMs trained with extensive code data to generate plausible test cases.
With the guidance of test cases, LLMs can comprehensively understand requirements and determine related details (\eg input-output formats, boundary inputs, outliers), thus generating more correct programs.

\vspace{-0.2cm}
\begin{tcolorbox}[size=title]
\textbf{Answer to RQ5:} We explore the other four designs for \method and compare them to our designs. Results on three benchmarks show the superiority of our design.
\end{tcolorbox}
\vspace{-0.2cm}

\vspace{-0.3cm}
\section{Discussion}
\label{sec:discussion}

\subsection{\method \textit{vs.} CoT prompting}

Our guided code generation is similar to Chain-of-Thought (CoT) prompting. Both approaches ask LLMs to first generate an intermediate result and then output the final code. 
The intermediate result in CoT prompting is a series of natural language steps describing how to write code step by step. In contrast, \method leverages some software artifacts (\eg test cases) as the intermediate result. 

We argue that our guided code generation is superior to the CoT in code generation. 
Table \ref{tab:RQ1} shows the comparison results between \method and CoT prompting.
CoT prompting achieves slight improvements over few-shot prompting and is even worse than zero-shot prompting. We inspect some failed samples and summarize the main reason. We find that CoTs describe how to write code in a series of steps almost at the same level as code. 
The LLMs for source code are mainly pre-trained with code data and are relatively weak in natural language generation. The generated CoTs often contain ambiguities or errors and negatively affect the subsequent code generation. Similar findings can be found in the original paper of CoT prompting \cite{Chain-of-thought}. 
Compared to CoT prompting, \method uses a software artifact (\ie test cases) as intermediate preliminaries. Compared to natural languages, test cases are more suitable to clarify requirements and contain fewer ambiguities. Besides, test cases are common in real-world code files, and LLMs have abilities to generate plausible test cases. 
Thus, \method is different from CoT prompting and is more promising than CoT prompting in code generation.

\vspace{-0.2cm}
\subsection{\method \textit{vs.} Rank Techniques}

Some recent studies \cite{CodeT,FaultCodeRanker} propose \textit{rank techniques} to improve the performance of LLMs on code generation. Given a requirement, they first sample many programs from LLMs and then use test cases or neural networks 
to rerank sampled programs. 

In this paper, we do not directly compare our approach to rank techniques. The reason is that \method and rank techniques have different focuses and they are complementary.
Our work is a new prompting technique that improves the accuracy of LLMs in code generation. Rank techniques do not care about LLMs and aim to select the best one from LLMs' multiple outputs. In practice, users can use \method to generate many programs and then use rank techniques to pick a final output. Thus, we omit them in experiments.

\vspace{-0.2cm}
\subsection{Threats to Validity}
\label{sec:discussion:threat}

There are two main threats to the validity of our work.

\textbf{The generalizability of experimental results.}
To mitigate this threat, we carefully select the experimental datasets, metrics, and baselines. 
Following previous studies \cite{MBXP, CodeT}, we pick three representative code generation benchmarks. They are collected from real-world software projects and cover three popular programming languages (\ie Python, Java, and JavaScript). For evaluation metrics, we select a widely used metric - Pass@$k$ ($k=1,3,5$). Pass@$k$ is an execution-based metric that utilizes test cases to check the correctness of programs.
We select existing prompting techniques and retrieval-based models as comparison baselines.
We pick three representative LLMs as base models \cite{Codex, CodeGeeX, CodeGen, InCoder}, which scale from 6B to 13B. 
We apply \method and baselines to base models and evaluate their performance on three datasets using Pass@k. To ensure fairness, we run each approach three times and report the average results.

\textbf{The impact of retrieved programs.}
The retrieved programs are important elements in \method. 
Intuitively, when retrieved programs are less relevant to input requirements, the performance of our approach may suffer. To address this threat, we have two thoughts. (1) A large-scale study on 13.2 million real code files found the proportion of reused code is up to 80\% \cite{Study_Code_Reuse}. Thus, we believe that it is quite possible to retrieve similar programs in real development scenarios. (2) Even if retrieved programs are less relevant to input requirements, \method degrades to few-shot prompting at worst. In most cases, \method is superior to few-shot prompting.

\vspace{-0.3cm}
\section{Related Work}
\label{sec:related_work}

\textbf{Large Language Models (LLMs) for Code Generation} are large-scale neural networks pre-trained on a large corpus of natural language and programming language. With the development of LLM research, current Code LLMs can be divided into two categories: standard language models and instruction-tuned models.

\textbf{Standard Language models} are pre-trained on the raw corpus with the next-token prediction. They can continually complete the given context, which makes them useful in tasks like code completion and code generation.
With the success of GPT series \cite{GPT-series, GPT-2, GPT-3} in NLP, 
OpenAI adapts similar idea into the domain of source code, and fine-tunes GPT models on code to produce closed-source Codex \cite{Codex}.
There are multiple open-source attempts to replicate its success, \eg CodeParrot \cite{CodeParrot}, CodeGen \cite{CodeGen}, CodeGeeX \cite{CodeGeeX}, InCoder \cite{InCoder}, StarCoder \cite{StarCoder} and CodeT5+ \cite{CodeT5+}.

\textbf{Instruction-tuned models} are models fine-tuned using instruction tuning \cite{instruct-tuning}. Instruction tuning helps models to follow users' instructions.
OpenAI's ChatGPT \cite{ChatGPT} is trained by Reinforcement Learning with Human Feedback (RLHF) \cite{RLHF}, making it capable of both natural language tasks and programming tasks.
Due to its enormous influence and closed-sourceness, many researchers try to create open-source ChatGPT alternatives using instruction tuning and its variants.
Alpaca \cite{alpaca} is LLaMA \cite{LLaMA} fine-tuned using self-instruct \cite{self-instruct} and ChatGPT feedback.
Code Alpaca \cite{code-alpaca} is LLaMA fine-tuned using self-instruct and ChatGPT feedback with more programming-focused instructions.
WizardCoder \cite{WizardCoder} is StarCoder \cite{StarCoder} fine-tuned using Evol-Instruct \cite{WizardLM} and ChatGPT feedback with Code Alpaca's dataset as seed dataset.
InstructCodeT5+ \cite{CodeT5+} is CodeT5+ \cite{CodeT5+} fine-tuned on Code Alpaca's dataset.

\textbf{Prompting Techniques.}
LLMs are too large to fine-tune, so researchers need to find a new way to adapt the LLMs on the downstream tasks.
\textit{Prompting techniques} are a popular approach to leverage LLMs to generate code by inputting a special prompt.

Early, researchers proposed zero-shot prompting and few-shot prompting. Zero-shot prompting concatenates a task instruction (\eg \texttt{please generate a program based on the requirement}) and a requirement together to make the prompt. Based on the zero-shot prompting, few-shot prompting further adds several $\left<\right.$requirement, code$\left.\right>$ pairs to the prompts, so that LLMs can learn code generation from given examples.
Chain-of-Thought (CoT) prompting \cite{Chain-of-thought} is a recently proposed prompting technique. 
CoT asks LLMs first to generate CoTs (\ie intermediate natural language reasoning steps) and then output the final code.
It allows LLMs to first design a solving process that leads to the code. CoT has achieved the SOTA results in natural language generation and sparked lots of follow-up research, such as self-consistency prompting \cite{Self-Consistency}, least-to-most prompting \cite{Least-to-Most}. But these prompting techniques are designed for natural language generation and bring slight improvements in code generation.

\vspace{-0.3cm}
\section{Conclusion and Future Work}
\label{sec:conclusion}

We propose a new prompting technique named \method to improve the performance of LLMs on code generation.
\method designs two novel techniques (\ie guided code generation and example retrieval) to help LLMs understand requirements and implement programs. 
Guided code generation asks LLMs to output an intermediate preliminary (\eg test cases) before generating programs. The preliminary helps LLMs understand requirements and guides the next code generation.
Example retrieval selects similar programs as examples, which provide many reusable elements for program implementation. 
We apply \method to three LLMs and conduct experiments on three benchmarks. Results show that \method significantly outperforms the SOTA baselines. 

In the future, we will explore how to improve the usability of LLMs in code generation. For example, how to teach LLMs to use unseen frameworks without re-training.


\normalem
\bibliographystyle{ACM-Reference-Format}
\bibliography{reference}


\begin{thebibliography}{52}


\ifx \showCODEN    \undefined \def \showCODEN     #1{\unskip}     \fi
\ifx \showDOI      \undefined \def \showDOI       #1{#1}\fi
\ifx \showISBNx    \undefined \def \showISBNx     #1{\unskip}     \fi
\ifx \showISBNxiii \undefined \def \showISBNxiii  #1{\unskip}     \fi
\ifx \showISSN     \undefined \def \showISSN      #1{\unskip}     \fi
\ifx \showLCCN     \undefined \def \showLCCN      #1{\unskip}     \fi
\ifx \shownote     \undefined \def \shownote      #1{#1}          \fi
\ifx \showarticletitle \undefined \def \showarticletitle #1{#1}   \fi
\ifx \showURL      \undefined \def \showURL       {\relax}        \fi
\providecommand\bibfield[2]{#2}
\providecommand\bibinfo[2]{#2}
\providecommand\natexlab[1]{#1}
\providecommand\showeprint[2][]{arXiv:#2}

\bibitem[Cod(2022a)]%
        {CodeGeeX}
 \bibinfo{year}{2022}\natexlab{a}.
\newblock \bibinfo{title}{CodeGeeX}.
\newblock
  \bibinfo{howpublished}{\url{https://models.aminer.cn/codegeex/zh-CN}}.
\newblock


\bibitem[Cod(2022b)]%
        {CodeGeeX-web}
 \bibinfo{year}{2022}\natexlab{b}.
\newblock \bibinfo{title}{CodeGeeX}.
\newblock
  \bibinfo{howpublished}{\url{https://models.aminer.cn/codegeex/blog/index.html}}.
\newblock


\bibitem[Cod(2022c)]%
        {CodeParrot}
 \bibinfo{year}{2022}\natexlab{c}.
\newblock \bibinfo{title}{CodeParrot}.
\newblock
  \bibinfo{howpublished}{\url{https://huggingface.co/codeparrot/codeparrot}}.
\newblock


\bibitem[Git(2022)]%
        {GitHub}
 \bibinfo{year}{2022}\natexlab{}.
\newblock \bibinfo{title}{GitHub}.
\newblock \bibinfo{howpublished}{\url{https://github.com/}}.
\newblock


\bibitem[Luc(2022)]%
        {Lucene}
 \bibinfo{year}{2022}\natexlab{}.
\newblock \bibinfo{title}{Lucene}.
\newblock \bibinfo{howpublished}{\url{https://lucene.apache.org/}}.
\newblock


\bibitem[tre(2022)]%
        {tree-sitter}
 \bibinfo{year}{2022}\natexlab{}.
\newblock \bibinfo{title}{tree-sitter}.
\newblock
  \bibinfo{howpublished}{\url{https://tree-sitter.github.io/tree-sitter/}}.
\newblock


\bibitem[Ahmad et~al\mbox{.}(2021)]%
        {PLBART}
\bibfield{author}{\bibinfo{person}{Wasi Ahmad}, \bibinfo{person}{Saikat
  Chakraborty}, \bibinfo{person}{Baishakhi Ray}, {and} \bibinfo{person}{Kai-Wei
  Chang}.} \bibinfo{year}{2021}\natexlab{}.
\newblock \showarticletitle{Unified Pre-training for Program Understanding and
  Generation}. In \bibinfo{booktitle}{\emph{Proceedings of the 2021 Conference
  of the North American Chapter of the Association for Computational
  Linguistics: Human Language Technologies}}. \bibinfo{publisher}{Association
  for Computational Linguistics}, \bibinfo{address}{Online},
  \bibinfo{pages}{2655--2668}.
\newblock
\urldef\tempurl%
\url{https://doi.org/10.18653/v1/2021.naacl-main.211}
\showDOI{\tempurl}


\bibitem[Athiwaratkun et~al\mbox{.}(2022)]%
        {MBXP}
\bibfield{author}{\bibinfo{person}{Ben Athiwaratkun},
  \bibinfo{person}{Sanjay~Krishna Gouda}, \bibinfo{person}{Zijian Wang},
  \bibinfo{person}{Xiaopeng Li}, \bibinfo{person}{Yuchen Tian},
  \bibinfo{person}{Ming Tan}, \bibinfo{person}{Wasi~Uddin Ahmad},
  \bibinfo{person}{Shiqi Wang}, \bibinfo{person}{Qing Sun},
  \bibinfo{person}{Mingyue Shang}, {et~al\mbox{.}}}
  \bibinfo{year}{2022}\natexlab{}.
\newblock \showarticletitle{Multi-lingual Evaluation of Code Generation
  Models}.
\newblock \bibinfo{journal}{\emph{arXiv preprint arXiv:2210.14868}}
  (\bibinfo{year}{2022}).
\newblock


\bibitem[Austin et~al\mbox{.}(2021)]%
        {MBPP}
\bibfield{author}{\bibinfo{person}{Jacob Austin}, \bibinfo{person}{Augustus
  Odena}, \bibinfo{person}{Maxwell Nye}, \bibinfo{person}{Maarten Bosma},
  \bibinfo{person}{Henryk Michalewski}, \bibinfo{person}{David Dohan},
  \bibinfo{person}{Ellen Jiang}, \bibinfo{person}{Carrie Cai},
  \bibinfo{person}{Michael Terry}, \bibinfo{person}{Quoc Le}, {et~al\mbox{.}}}
  \bibinfo{year}{2021}\natexlab{}.
\newblock \showarticletitle{Program synthesis with large language models}.
\newblock \bibinfo{journal}{\emph{arXiv preprint arXiv:2108.07732}}
  (\bibinfo{year}{2021}).
\newblock


\bibitem[Beck(2003)]%
        {Test_Driven_Development}
\bibfield{author}{\bibinfo{person}{Kent Beck}.}
  \bibinfo{year}{2003}\natexlab{}.
\newblock \bibinfo{booktitle}{\emph{Test-driven development: by example}}.
\newblock \bibinfo{publisher}{Addison-Wesley Professional}.
\newblock


\bibitem[Brown et~al\mbox{.}(2020)]%
        {GPT-3}
\bibfield{author}{\bibinfo{person}{Tom Brown}, \bibinfo{person}{Benjamin Mann},
  \bibinfo{person}{Nick Ryder}, \bibinfo{person}{Melanie Subbiah},
  \bibinfo{person}{Jared~D Kaplan}, \bibinfo{person}{Prafulla Dhariwal},
  \bibinfo{person}{Arvind Neelakantan}, \bibinfo{person}{Pranav Shyam},
  \bibinfo{person}{Girish Sastry}, \bibinfo{person}{Amanda Askell},
  {et~al\mbox{.}}} \bibinfo{year}{2020}\natexlab{}.
\newblock \showarticletitle{Language models are few-shot learners}.
\newblock \bibinfo{journal}{\emph{Advances in neural information processing
  systems}}  \bibinfo{volume}{33} (\bibinfo{year}{2020}),
  \bibinfo{pages}{1877--1901}.
\newblock


\bibitem[Chaudhary(2023)]%
        {code-alpaca}
\bibfield{author}{\bibinfo{person}{Sahil Chaudhary}.}
  \bibinfo{year}{2023}\natexlab{}.
\newblock \bibinfo{title}{Code Alpaca: An Instruction-following LLaMA model for
  code generation}.
\newblock
  \bibinfo{howpublished}{\url{https://github.com/sahil280114/codealpaca}}.
\newblock


\bibitem[Chen et~al\mbox{.}(2022)]%
        {CodeT}
\bibfield{author}{\bibinfo{person}{Bei Chen}, \bibinfo{person}{Fengji Zhang},
  \bibinfo{person}{Anh Nguyen}, \bibinfo{person}{Daoguang Zan},
  \bibinfo{person}{Zeqi Lin}, \bibinfo{person}{Jian-Guang Lou}, {and}
  \bibinfo{person}{Weizhu Chen}.} \bibinfo{year}{2022}\natexlab{}.
\newblock \showarticletitle{Codet: Code generation with generated tests}.
\newblock \bibinfo{journal}{\emph{arXiv preprint arXiv:2207.10397}}
  (\bibinfo{year}{2022}).
\newblock


\bibitem[Chen et~al\mbox{.}(2021)]%
        {Codex}
\bibfield{author}{\bibinfo{person}{Mark Chen}, \bibinfo{person}{Jerry Tworek},
  \bibinfo{person}{Heewoo Jun}, \bibinfo{person}{Qiming Yuan},
  \bibinfo{person}{Henrique Ponde de~Oliveira Pinto}, \bibinfo{person}{Jared
  Kaplan}, \bibinfo{person}{Harri Edwards}, \bibinfo{person}{Yuri Burda},
  \bibinfo{person}{Nicholas Joseph}, \bibinfo{person}{Greg Brockman},
  {et~al\mbox{.}}} \bibinfo{year}{2021}\natexlab{}.
\newblock \showarticletitle{Evaluating large language models trained on code}.
\newblock \bibinfo{journal}{\emph{arXiv preprint arXiv:2107.03374}}
  (\bibinfo{year}{2021}).
\newblock


\bibitem[Fried et~al\mbox{.}(2022)]%
        {InCoder}
\bibfield{author}{\bibinfo{person}{Daniel Fried}, \bibinfo{person}{Armen
  Aghajanyan}, \bibinfo{person}{Jessy Lin}, \bibinfo{person}{Sida Wang},
  \bibinfo{person}{Eric Wallace}, \bibinfo{person}{Freda Shi},
  \bibinfo{person}{Ruiqi Zhong}, \bibinfo{person}{Wen-tau Yih},
  \bibinfo{person}{Luke Zettlemoyer}, {and} \bibinfo{person}{Mike Lewis}.}
  \bibinfo{year}{2022}\natexlab{}.
\newblock \showarticletitle{Incoder: A generative model for code infilling and
  synthesis}.
\newblock \bibinfo{journal}{\emph{arXiv preprint arXiv:2204.05999}}
  (\bibinfo{year}{2022}).
\newblock


\bibitem[Hao et~al\mbox{.}(2022)]%
        {AixBench}
\bibfield{author}{\bibinfo{person}{Yiyang Hao}, \bibinfo{person}{Ge Li},
  \bibinfo{person}{Yongqiang Liu}, \bibinfo{person}{Xiaowei Miao},
  \bibinfo{person}{He Zong}, \bibinfo{person}{Siyuan Jiang},
  \bibinfo{person}{Yang Liu}, {and} \bibinfo{person}{He Wei}.}
  \bibinfo{year}{2022}\natexlab{}.
\newblock \showarticletitle{AixBench: A Code Generation Benchmark Dataset}.
\newblock \bibinfo{journal}{\emph{arXiv preprint arXiv:2206.13179}}
  (\bibinfo{year}{2022}).
\newblock


\bibitem[Hashimoto et~al\mbox{.}(2018)]%
        {Retrieve_edit_code_generation}
\bibfield{author}{\bibinfo{person}{Tatsunori~B Hashimoto},
  \bibinfo{person}{Kelvin Guu}, \bibinfo{person}{Yonatan Oren}, {and}
  \bibinfo{person}{Percy~S Liang}.} \bibinfo{year}{2018}\natexlab{}.
\newblock \showarticletitle{A retrieve-and-edit framework for predicting
  structured outputs}.
\newblock \bibinfo{journal}{\emph{Advances in Neural Information Processing
  Systems}}  \bibinfo{volume}{31} (\bibinfo{year}{2018}).
\newblock


\bibitem[Hendrycks et~al\mbox{.}(2021)]%
        {APPS}
\bibfield{author}{\bibinfo{person}{Dan Hendrycks}, \bibinfo{person}{Steven
  Basart}, \bibinfo{person}{Saurav Kadavath}, \bibinfo{person}{Mantas Mazeika},
  \bibinfo{person}{Akul Arora}, \bibinfo{person}{Ethan Guo},
  \bibinfo{person}{Collin Burns}, \bibinfo{person}{Samir Puranik},
  \bibinfo{person}{Horace He}, \bibinfo{person}{Dawn Song}, {et~al\mbox{.}}}
  \bibinfo{year}{2021}\natexlab{}.
\newblock In \bibinfo{booktitle}{\emph{Thirty-fifth Conference on Neural
  Information Processing Systems Datasets and Benchmarks Track (Round 2)}}.
\newblock


\bibitem[Holtzman et~al\mbox{.}(2020)]%
        {nucleus_sampling}
\bibfield{author}{\bibinfo{person}{Ari Holtzman}, \bibinfo{person}{Jan Buys},
  \bibinfo{person}{Li Du}, \bibinfo{person}{Maxwell Forbes}, {and}
  \bibinfo{person}{Yejin Choi}.} \bibinfo{year}{2020}\natexlab{}.
\newblock \showarticletitle{The Curious Case of Neural Text Degeneration}. In
  \bibinfo{booktitle}{\emph{8th International Conference on Learning
  Representations, {ICLR} 2020, Addis Ababa, Ethiopia, April 26-30, 2020}}.
  \bibinfo{publisher}{OpenReview.net}.
\newblock
\urldef\tempurl%
\url{https://openreview.net/forum?id=rygGQyrFvH}
\showURL{%
\tempurl}


\bibitem[Inala et~al\mbox{.}(2022)]%
        {FaultCodeRanker}
\bibfield{author}{\bibinfo{person}{Jeevana~Priya Inala},
  \bibinfo{person}{Chenglong Wang}, \bibinfo{person}{Mei Yang},
  \bibinfo{person}{Andr{\'{e}}s Codas}, \bibinfo{person}{Mark
  Encarnaci{\'{o}}n}, \bibinfo{person}{Shuvendu~K. Lahiri},
  \bibinfo{person}{Madanlal Musuvathi}, {and} \bibinfo{person}{Jianfeng Gao}.}
  \bibinfo{year}{2022}\natexlab{}.
\newblock \showarticletitle{Fault-Aware Neural Code Rankers}. In
  \bibinfo{booktitle}{\emph{NeurIPS}}.
\newblock
\urldef\tempurl%
\url{http://papers.nips.cc/paper\_files/paper/2022/hash/5762c579d09811b7639be2389b3d07be-Abstract-Conference.html}
\showURL{%
\tempurl}


\bibitem[Jain et~al\mbox{.}(2022)]%
        {Jigsaw}
\bibfield{author}{\bibinfo{person}{Naman Jain}, \bibinfo{person}{Skanda
  Vaidyanath}, \bibinfo{person}{Arun Iyer}, \bibinfo{person}{Nagarajan
  Natarajan}, \bibinfo{person}{Suresh Parthasarathy}, \bibinfo{person}{Sriram
  Rajamani}, {and} \bibinfo{person}{Rahul Sharma}.}
  \bibinfo{year}{2022}\natexlab{}.
\newblock \showarticletitle{Jigsaw: Large language models meet program
  synthesis}. In \bibinfo{booktitle}{\emph{Proceedings of the 44th
  International Conference on Software Engineering}}.
  \bibinfo{pages}{1219--1231}.
\newblock


\bibitem[Li et~al\mbox{.}(2023b)]%
        {TiP}
\bibfield{author}{\bibinfo{person}{Jia Li}, \bibinfo{person}{Ge Li},
  \bibinfo{person}{Yongmin Li}, {and} \bibinfo{person}{Zhi Jin}.}
  \bibinfo{year}{2023}\natexlab{b}.
\newblock \showarticletitle{Enabling Programming Thinking in Large Language
  Models Toward Code Generation}.
\newblock \bibinfo{journal}{\emph{CoRR}}  \bibinfo{volume}{abs/2305.06599}
  (\bibinfo{year}{2023}).
\newblock
\urldef\tempurl%
\url{https://doi.org/10.48550/arXiv.2305.06599}
\showDOI{\tempurl}
\showeprint[arXiv]{2305.06599}


\bibitem[Li et~al\mbox{.}(2023d)]%
        {CodeEditor}
\bibfield{author}{\bibinfo{person}{Jia Li}, \bibinfo{person}{Ge Li},
  \bibinfo{person}{Zhuo Li}, \bibinfo{person}{Zhi Jin}, \bibinfo{person}{Xing
  Hu}, \bibinfo{person}{Kechi Zhang}, {and} \bibinfo{person}{Zhiyi Fu}.}
  \bibinfo{year}{2023}\natexlab{d}.
\newblock \showarticletitle{CodeEditor: Learning to Edit Source Code with
  Pre-Trained Models}.
\newblock \bibinfo{journal}{\emph{ACM Trans. Softw. Eng. Methodol.}}
  (\bibinfo{date}{may} \bibinfo{year}{2023}).
\newblock
\showISSN{1049-331X}
\urldef\tempurl%
\url{https://doi.org/10.1145/3597207}
\showDOI{\tempurl}
\newblock
\shownote{Just Accepted}.


\bibitem[Li et~al\mbox{.}(2021)]%
        {EditSum}
\bibfield{author}{\bibinfo{person}{Jia Li}, \bibinfo{person}{Yongmin Li},
  \bibinfo{person}{Ge Li}, \bibinfo{person}{Xing Hu}, \bibinfo{person}{Xin
  Xia}, {and} \bibinfo{person}{Zhi Jin}.} \bibinfo{year}{2021}\natexlab{}.
\newblock \showarticletitle{Editsum: A retrieve-and-edit framework for source
  code summarization}. In \bibinfo{booktitle}{\emph{2021 36th IEEE/ACM
  International Conference on Automated Software Engineering (ASE)}}. IEEE,
  \bibinfo{pages}{155--166}.
\newblock


\bibitem[Li et~al\mbox{.}(2023c)]%
        {SkCoder}
\bibfield{author}{\bibinfo{person}{Jia Li}, \bibinfo{person}{Yongmin Li},
  \bibinfo{person}{Ge Li}, \bibinfo{person}{Zhi Jin}, \bibinfo{person}{Yiyang
  Hao}, {and} \bibinfo{person}{Xing Hu}.} \bibinfo{year}{2023}\natexlab{c}.
\newblock \showarticletitle{SkCoder: {A} Sketch-based Approach for Automatic
  Code Generation}. In \bibinfo{booktitle}{\emph{45th {IEEE/ACM} International
  Conference on Software Engineering, {ICSE} 2023, Melbourne, Australia, May
  14-20, 2023}}. \bibinfo{publisher}{{IEEE}}, \bibinfo{pages}{2124--2135}.
\newblock
\urldef\tempurl%
\url{https://doi.org/10.1109/ICSE48619.2023.00179}
\showDOI{\tempurl}


\bibitem[Li et~al\mbox{.}(2023a)]%
        {StarCoder}
\bibfield{author}{\bibinfo{person}{Raymond Li}, \bibinfo{person}{Loubna~Ben
  Allal}, \bibinfo{person}{Yangtian Zi}, \bibinfo{person}{Niklas Muennighoff},
  \bibinfo{person}{Denis Kocetkov}, \bibinfo{person}{Chenghao Mou},
  \bibinfo{person}{Marc Marone}, \bibinfo{person}{Christopher Akiki},
  \bibinfo{person}{Jia Li}, \bibinfo{person}{Jenny Chim}, {et~al\mbox{.}}}
  \bibinfo{year}{2023}\natexlab{a}.
\newblock \showarticletitle{StarCoder: may the source be with you!}
\newblock \bibinfo{journal}{\emph{arXiv preprint arXiv:2305.06161}}
  (\bibinfo{year}{2023}).
\newblock


\bibitem[Li et~al\mbox{.}(2022)]%
        {AlphaCode}
\bibfield{author}{\bibinfo{person}{Yujia Li}, \bibinfo{person}{David Choi},
  \bibinfo{person}{Junyoung Chung}, \bibinfo{person}{Nate Kushman},
  \bibinfo{person}{Julian Schrittwieser}, \bibinfo{person}{R{\'e}mi Leblond},
  \bibinfo{person}{Tom Eccles}, \bibinfo{person}{James Keeling},
  \bibinfo{person}{Felix Gimeno}, \bibinfo{person}{Agustin Dal~Lago},
  {et~al\mbox{.}}} \bibinfo{year}{2022}\natexlab{}.
\newblock \showarticletitle{Competition-level code generation with alphacode}.
\newblock \bibinfo{journal}{\emph{Science}} \bibinfo{volume}{378},
  \bibinfo{number}{6624} (\bibinfo{year}{2022}), \bibinfo{pages}{1092--1097}.
\newblock


\bibitem[LIN(2004)]%
        {ROUGE}
\bibfield{author}{\bibinfo{person}{CY LIN}.} \bibinfo{year}{2004}\natexlab{}.
\newblock \showarticletitle{Rouge: A package for automatic evaluation of
  summaries}. In \bibinfo{booktitle}{\emph{Text Summarization Branches Out:
  Proceedings of the ACL-04 Workshop, Barcelona, Spain}}.
  \bibinfo{pages}{74--81}.
\newblock


\bibitem[Lu et~al\mbox{.}(2022)]%
        {ReAcc}
\bibfield{author}{\bibinfo{person}{Shuai Lu}, \bibinfo{person}{Nan Duan},
  \bibinfo{person}{Hojae Han}, \bibinfo{person}{Daya Guo},
  \bibinfo{person}{Seung{-}won Hwang}, {and} \bibinfo{person}{Alexey
  Svyatkovskiy}.} \bibinfo{year}{2022}\natexlab{}.
\newblock \showarticletitle{ReACC: {A} Retrieval-Augmented Code Completion
  Framework}. In \bibinfo{booktitle}{\emph{Proceedings of the 60th Annual
  Meeting of the Association for Computational Linguistics (Volume 1: Long
  Papers), {ACL} 2022, Dublin, Ireland, May 22-27, 2022}},
  \bibfield{editor}{\bibinfo{person}{Smaranda Muresan},
  \bibinfo{person}{Preslav Nakov}, {and} \bibinfo{person}{Aline Villavicencio}}
  (Eds.). \bibinfo{publisher}{Association for Computational Linguistics},
  \bibinfo{pages}{6227--6240}.
\newblock
\urldef\tempurl%
\url{https://doi.org/10.18653/v1/2022.acl-long.431}
\showDOI{\tempurl}


\bibitem[Luo et~al\mbox{.}(2023)]%
        {WizardCoder}
\bibfield{author}{\bibinfo{person}{Ziyang Luo}, \bibinfo{person}{Can Xu},
  \bibinfo{person}{Pu Zhao}, \bibinfo{person}{Qingfeng Sun},
  \bibinfo{person}{Xiubo Geng}, \bibinfo{person}{Wenxiang Hu},
  \bibinfo{person}{Chongyang Tao}, \bibinfo{person}{Jing Ma},
  \bibinfo{person}{Qingwei Lin}, {and} \bibinfo{person}{Daxin Jiang}.}
  \bibinfo{year}{2023}\natexlab{}.
\newblock \showarticletitle{WizardCoder: Empowering Code Large Language Models
  with Evol-Instruct}.
\newblock \bibinfo{journal}{\emph{arXiv preprint arXiv:2306.08568}}
  (\bibinfo{year}{2023}).
\newblock


\bibitem[Mockus(2007)]%
        {Study_Code_Reuse}
\bibfield{author}{\bibinfo{person}{Audris Mockus}.}
  \bibinfo{year}{2007}\natexlab{}.
\newblock \showarticletitle{Large-scale code reuse in open source software}. In
  \bibinfo{booktitle}{\emph{First International Workshop on Emerging Trends in
  FLOSS Research and Development (FLOSS'07: ICSE Workshops 2007)}}. IEEE,
  \bibinfo{pages}{7--7}.
\newblock


\bibitem[Nijkamp et~al\mbox{.}(2022)]%
        {CodeGen}
\bibfield{author}{\bibinfo{person}{Erik Nijkamp}, \bibinfo{person}{Bo Pang},
  \bibinfo{person}{Hiroaki Hayashi}, \bibinfo{person}{Lifu Tu},
  \bibinfo{person}{Huan Wang}, \bibinfo{person}{Yingbo Zhou},
  \bibinfo{person}{Silvio Savarese}, {and} \bibinfo{person}{Caiming Xiong}.}
  \bibinfo{year}{2022}\natexlab{}.
\newblock \showarticletitle{A conversational paradigm for program synthesis}.
\newblock \bibinfo{journal}{\emph{arXiv preprint arXiv:2203.13474}}
  (\bibinfo{year}{2022}).
\newblock


\bibitem[OpenAI(2022)]%
        {ChatGPT}
\bibfield{author}{\bibinfo{person}{OpenAI}.} \bibinfo{year}{2022}\natexlab{}.
\newblock \bibinfo{title}{ChatGPT}.
\newblock \bibinfo{howpublished}{\url{https://openai.com/blog/chatgpt}}.
\newblock


\bibitem[Ouyang et~al\mbox{.}(2022)]%
        {RLHF}
\bibfield{author}{\bibinfo{person}{Long Ouyang}, \bibinfo{person}{Jeffrey Wu},
  \bibinfo{person}{Xu Jiang}, \bibinfo{person}{Diogo Almeida},
  \bibinfo{person}{Carroll Wainwright}, \bibinfo{person}{Pamela Mishkin},
  \bibinfo{person}{Chong Zhang}, \bibinfo{person}{Sandhini Agarwal},
  \bibinfo{person}{Katarina Slama}, \bibinfo{person}{Alex Ray},
  {et~al\mbox{.}}} \bibinfo{year}{2022}\natexlab{}.
\newblock \showarticletitle{Training language models to follow instructions
  with human feedback}.
\newblock \bibinfo{journal}{\emph{Advances in Neural Information Processing
  Systems}}  \bibinfo{volume}{35} (\bibinfo{year}{2022}),
  \bibinfo{pages}{27730--27744}.
\newblock


\bibitem[Papineni et~al\mbox{.}(2002)]%
        {BLEU}
\bibfield{author}{\bibinfo{person}{Kishore Papineni}, \bibinfo{person}{Salim
  Roukos}, \bibinfo{person}{Todd Ward}, {and} \bibinfo{person}{Wei-Jing Zhu}.}
  \bibinfo{year}{2002}\natexlab{}.
\newblock \showarticletitle{Bleu: a method for automatic evaluation of machine
  translation}. In \bibinfo{booktitle}{\emph{Proceedings of the 40th annual
  meeting of the Association for Computational Linguistics}}.
  \bibinfo{pages}{311--318}.
\newblock


\bibitem[Parvez et~al\mbox{.}(2021)]%
        {REDCODER}
\bibfield{author}{\bibinfo{person}{Md.~Rizwan Parvez},
  \bibinfo{person}{Wasi~Uddin Ahmad}, \bibinfo{person}{Saikat Chakraborty},
  \bibinfo{person}{Baishakhi Ray}, {and} \bibinfo{person}{Kai{-}Wei Chang}.}
  \bibinfo{year}{2021}\natexlab{}.
\newblock \showarticletitle{Retrieval Augmented Code Generation and
  Summarization}. In \bibinfo{booktitle}{\emph{Findings of the Association for
  Computational Linguistics: {EMNLP} 2021, Virtual Event / Punta Cana,
  Dominican Republic, 16-20 November, 2021}},
  \bibfield{editor}{\bibinfo{person}{Marie{-}Francine Moens},
  \bibinfo{person}{Xuanjing Huang}, \bibinfo{person}{Lucia Specia}, {and}
  \bibinfo{person}{Scott~Wen{-}tau Yih}} (Eds.).
  \bibinfo{publisher}{Association for Computational Linguistics},
  \bibinfo{pages}{2719--2734}.
\newblock
\urldef\tempurl%
\url{https://doi.org/10.18653/v1/2021.findings-emnlp.232}
\showDOI{\tempurl}


\bibitem[Radford et~al\mbox{.}(2018)]%
        {GPT-series}
\bibfield{author}{\bibinfo{person}{Alec Radford}, \bibinfo{person}{Karthik
  Narasimhan}, \bibinfo{person}{Tim Salimans}, \bibinfo{person}{Ilya
  Sutskever}, {et~al\mbox{.}}} \bibinfo{year}{2018}\natexlab{}.
\newblock \showarticletitle{Improving language understanding by generative
  pre-training}.
\newblock  (\bibinfo{year}{2018}).
\newblock


\bibitem[Radford et~al\mbox{.}(2019)]%
        {GPT-2}
\bibfield{author}{\bibinfo{person}{Alec Radford}, \bibinfo{person}{Jeffrey Wu},
  \bibinfo{person}{Rewon Child}, \bibinfo{person}{David Luan},
  \bibinfo{person}{Dario Amodei}, \bibinfo{person}{Ilya Sutskever},
  {et~al\mbox{.}}} \bibinfo{year}{2019}\natexlab{}.
\newblock \showarticletitle{Language models are unsupervised multitask
  learners}.
\newblock \bibinfo{journal}{\emph{OpenAI blog}} \bibinfo{volume}{1},
  \bibinfo{number}{8} (\bibinfo{year}{2019}), \bibinfo{pages}{9}.
\newblock


\bibitem[Reimers and Gurevych(2019)]%
        {SentenceBERT}
\bibfield{author}{\bibinfo{person}{Nils Reimers} {and} \bibinfo{person}{Iryna
  Gurevych}.} \bibinfo{year}{2019}\natexlab{}.
\newblock \showarticletitle{Sentence-BERT: Sentence Embeddings using Siamese
  BERT-Networks}. In \bibinfo{booktitle}{\emph{Proceedings of the 2019
  Conference on Empirical Methods in Natural Language Processing and the 9th
  International Joint Conference on Natural Language Processing, {EMNLP-IJCNLP}
  2019, Hong Kong, China, November 3-7, 2019}},
  \bibfield{editor}{\bibinfo{person}{Kentaro Inui}, \bibinfo{person}{Jing
  Jiang}, \bibinfo{person}{Vincent Ng}, {and} \bibinfo{person}{Xiaojun Wan}}
  (Eds.). \bibinfo{publisher}{Association for Computational Linguistics},
  \bibinfo{pages}{3980--3990}.
\newblock
\urldef\tempurl%
\url{https://doi.org/10.18653/v1/D19-1410}
\showDOI{\tempurl}


\bibitem[Robertson et~al\mbox{.}(2009)]%
        {BM25}
\bibfield{author}{\bibinfo{person}{Stephen Robertson}, \bibinfo{person}{Hugo
  Zaragoza}, {et~al\mbox{.}}} \bibinfo{year}{2009}\natexlab{}.
\newblock \showarticletitle{The probabilistic relevance framework: BM25 and
  beyond}.
\newblock \bibinfo{journal}{\emph{Foundations and Trends{\textregistered} in
  Information Retrieval}} \bibinfo{volume}{3}, \bibinfo{number}{4}
  (\bibinfo{year}{2009}), \bibinfo{pages}{333--389}.
\newblock


\bibitem[Taori et~al\mbox{.}(2023)]%
        {alpaca}
\bibfield{author}{\bibinfo{person}{Rohan Taori}, \bibinfo{person}{Ishaan
  Gulrajani}, \bibinfo{person}{Tianyi Zhang}, \bibinfo{person}{Yann Dubois},
  \bibinfo{person}{Xuechen Li}, \bibinfo{person}{Carlos Guestrin},
  \bibinfo{person}{Percy Liang}, {and} \bibinfo{person}{Tatsunori~B.
  Hashimoto}.} \bibinfo{year}{2023}\natexlab{}.
\newblock \bibinfo{title}{Stanford Alpaca: An Instruction-following LLaMA
  model}.
\newblock
  \bibinfo{howpublished}{\url{https://github.com/tatsu-lab/stanford_alpaca}}.
\newblock


\bibitem[Touvron et~al\mbox{.}(2023)]%
        {LLaMA}
\bibfield{author}{\bibinfo{person}{Hugo Touvron}, \bibinfo{person}{Thibaut
  Lavril}, \bibinfo{person}{Gautier Izacard}, \bibinfo{person}{Xavier
  Martinet}, \bibinfo{person}{Marie-Anne Lachaux},
  \bibinfo{person}{Timoth{\'e}e Lacroix}, \bibinfo{person}{Baptiste
  Rozi{\`e}re}, \bibinfo{person}{Naman Goyal}, \bibinfo{person}{Eric Hambro},
  \bibinfo{person}{Faisal Azhar}, \bibinfo{person}{Aurelien Rodriguez},
  \bibinfo{person}{Armand Joulin}, \bibinfo{person}{Edouard Grave}, {and}
  \bibinfo{person}{Guillaume Lample}.} \bibinfo{year}{2023}\natexlab{}.
\newblock \showarticletitle{LLaMA: Open and Efficient Foundation Language
  Models}.
\newblock \bibinfo{journal}{\emph{arXiv preprint arXiv:2302.13971}}
  (\bibinfo{year}{2023}).
\newblock


\bibitem[Wang et~al\mbox{.}(2023c)]%
        {Self-Consistency}
\bibfield{author}{\bibinfo{person}{Xuezhi Wang}, \bibinfo{person}{Jason Wei},
  \bibinfo{person}{Dale Schuurmans}, \bibinfo{person}{Quoc~V. Le},
  \bibinfo{person}{Ed~H. Chi}, \bibinfo{person}{Sharan Narang},
  \bibinfo{person}{Aakanksha Chowdhery}, {and} \bibinfo{person}{Denny Zhou}.}
  \bibinfo{year}{2023}\natexlab{c}.
\newblock \showarticletitle{Self-Consistency Improves Chain of Thought
  Reasoning in Language Models}. In \bibinfo{booktitle}{\emph{The Eleventh
  International Conference on Learning Representations, {ICLR} 2023, Kigali,
  Rwanda, May 1-5, 2023}}. \bibinfo{publisher}{OpenReview.net}.
\newblock
\urldef\tempurl%
\url{https://openreview.net/pdf?id=1PL1NIMMrw}
\showURL{%
\tempurl}


\bibitem[Wang et~al\mbox{.}(2023a)]%
        {self-instruct}
\bibfield{author}{\bibinfo{person}{Yizhong Wang}, \bibinfo{person}{Yeganeh
  Kordi}, \bibinfo{person}{Swaroop Mishra}, \bibinfo{person}{Alisa Liu},
  \bibinfo{person}{Noah~A. Smith}, \bibinfo{person}{Daniel Khashabi}, {and}
  \bibinfo{person}{Hannaneh Hajishirzi}.} \bibinfo{year}{2023}\natexlab{a}.
\newblock \showarticletitle{Self-Instruct: Aligning Language Models with
  Self-Generated Instructions}. In \bibinfo{booktitle}{\emph{Proceedings of the
  61st Annual Meeting of the Association for Computational Linguistics (Volume
  1: Long Papers)}}. \bibinfo{publisher}{Association for Computational
  Linguistics}, \bibinfo{address}{Toronto, Canada},
  \bibinfo{pages}{13484--13508}.
\newblock
\urldef\tempurl%
\url{https://aclanthology.org/2023.acl-long.754}
\showURL{%
\tempurl}


\bibitem[Wang et~al\mbox{.}(2023b)]%
        {CodeT5+}
\bibfield{author}{\bibinfo{person}{Yue Wang}, \bibinfo{person}{Hung Le},
  \bibinfo{person}{Akhilesh~Deepak Gotmare}, \bibinfo{person}{Nghi~DQ Bui},
  \bibinfo{person}{Junnan Li}, {and} \bibinfo{person}{Steven~CH Hoi}.}
  \bibinfo{year}{2023}\natexlab{b}.
\newblock \showarticletitle{Codet5+: Open code large language models for code
  understanding and generation}.
\newblock \bibinfo{journal}{\emph{arXiv preprint arXiv:2305.07922}}
  (\bibinfo{year}{2023}).
\newblock


\bibitem[Wang et~al\mbox{.}(2021)]%
        {CodeT5}
\bibfield{author}{\bibinfo{person}{Yue Wang}, \bibinfo{person}{Weishi Wang},
  \bibinfo{person}{Shafiq Joty}, {and} \bibinfo{person}{Steven~C.H. Hoi}.}
  \bibinfo{year}{2021}\natexlab{}.
\newblock \showarticletitle{{C}ode{T}5: Identifier-aware Unified Pre-trained
  Encoder-Decoder Models for Code Understanding and Generation}. In
  \bibinfo{booktitle}{\emph{Proceedings of the 2021 Conference on Empirical
  Methods in Natural Language Processing}}. \bibinfo{publisher}{Association for
  Computational Linguistics}, \bibinfo{address}{Online and Punta Cana,
  Dominican Republic}, \bibinfo{pages}{8696--8708}.
\newblock
\urldef\tempurl%
\url{https://doi.org/10.18653/v1/2021.emnlp-main.685}
\showDOI{\tempurl}


\bibitem[Wei et~al\mbox{.}(2020)]%
        {Re2Com}
\bibfield{author}{\bibinfo{person}{Bolin Wei}, \bibinfo{person}{Yongmin Li},
  \bibinfo{person}{Ge Li}, \bibinfo{person}{Xin Xia}, {and}
  \bibinfo{person}{Zhi Jin}.} \bibinfo{year}{2020}\natexlab{}.
\newblock \showarticletitle{Retrieve and refine: exemplar-based neural comment
  generation}. In \bibinfo{booktitle}{\emph{2020 35th IEEE/ACM International
  Conference on Automated Software Engineering (ASE)}}. IEEE,
  \bibinfo{pages}{349--360}.
\newblock


\bibitem[Wei et~al\mbox{.}(2022a)]%
        {instruct-tuning}
\bibfield{author}{\bibinfo{person}{Jason Wei}, \bibinfo{person}{Maarten Bosma},
  \bibinfo{person}{Vincent Zhao}, \bibinfo{person}{Kelvin Guu},
  \bibinfo{person}{Adams~Wei Yu}, \bibinfo{person}{Brian Lester},
  \bibinfo{person}{Nan Du}, \bibinfo{person}{Andrew~M. Dai}, {and}
  \bibinfo{person}{Quoc~V Le}.} \bibinfo{year}{2022}\natexlab{a}.
\newblock \showarticletitle{Finetuned Language Models are Zero-Shot Learners}.
  In \bibinfo{booktitle}{\emph{International Conference on Learning
  Representations}}.
\newblock
\urldef\tempurl%
\url{https://openreview.net/forum?id=gEZrGCozdqR}
\showURL{%
\tempurl}


\bibitem[Wei et~al\mbox{.}(2022b)]%
        {Chain-of-thought}
\bibfield{author}{\bibinfo{person}{Jason Wei}, \bibinfo{person}{Xuezhi Wang},
  \bibinfo{person}{Dale Schuurmans}, \bibinfo{person}{Maarten Bosma},
  \bibinfo{person}{Fei Xia}, \bibinfo{person}{Ed~H Chi},
  \bibinfo{person}{Quoc~V Le}, \bibinfo{person}{Denny Zhou}, {et~al\mbox{.}}}
  \bibinfo{year}{2022}\natexlab{b}.
\newblock \showarticletitle{Chain-of-Thought Prompting Elicits Reasoning in
  Large Language Models}. In \bibinfo{booktitle}{\emph{Advances in Neural
  Information Processing Systems}}.
\newblock


\bibitem[Xu et~al\mbox{.}(2023)]%
        {WizardLM}
\bibfield{author}{\bibinfo{person}{Can Xu}, \bibinfo{person}{Qingfeng Sun},
  \bibinfo{person}{Kai Zheng}, \bibinfo{person}{Xiubo Geng},
  \bibinfo{person}{Pu Zhao}, \bibinfo{person}{Jiazhan Feng},
  \bibinfo{person}{Chongyang Tao}, {and} \bibinfo{person}{Daxin Jiang}.}
  \bibinfo{year}{2023}\natexlab{}.
\newblock \showarticletitle{Wizardlm: Empowering large language models to
  follow complex instructions}.
\newblock \bibinfo{journal}{\emph{arXiv preprint arXiv:2304.12244}}
  (\bibinfo{year}{2023}).
\newblock


\bibitem[Zhao et~al\mbox{.}(2021)]%
        {PromptStudy}
\bibfield{author}{\bibinfo{person}{Zihao Zhao}, \bibinfo{person}{Eric Wallace},
  \bibinfo{person}{Shi Feng}, \bibinfo{person}{Dan Klein}, {and}
  \bibinfo{person}{Sameer Singh}.} \bibinfo{year}{2021}\natexlab{}.
\newblock \showarticletitle{Calibrate before use: Improving few-shot
  performance of language models}. In \bibinfo{booktitle}{\emph{International
  Conference on Machine Learning}}. PMLR, \bibinfo{pages}{12697--12706}.
\newblock


\bibitem[Zhou et~al\mbox{.}(2023)]%
        {Least-to-Most}
\bibfield{author}{\bibinfo{person}{Denny Zhou}, \bibinfo{person}{Nathanael
  Sch{\"{a}}rli}, \bibinfo{person}{Le Hou}, \bibinfo{person}{Jason Wei},
  \bibinfo{person}{Nathan Scales}, \bibinfo{person}{Xuezhi Wang},
  \bibinfo{person}{Dale Schuurmans}, \bibinfo{person}{Claire Cui},
  \bibinfo{person}{Olivier Bousquet}, \bibinfo{person}{Quoc~V. Le}, {and}
  \bibinfo{person}{Ed~H. Chi}.} \bibinfo{year}{2023}\natexlab{}.
\newblock \showarticletitle{Least-to-Most Prompting Enables Complex Reasoning
  in Large Language Models}. In \bibinfo{booktitle}{\emph{The Eleventh
  International Conference on Learning Representations, {ICLR} 2023, Kigali,
  Rwanda, May 1-5, 2023}}. \bibinfo{publisher}{OpenReview.net}.
\newblock
\urldef\tempurl%
\url{https://openreview.net/pdf?id=WZH7099tgfM}
\showURL{%
\tempurl}


\end{thebibliography}

\appendix

\end{document}